\documentclass[
prc,
preprint,
reprint,
superscriptaddress,
showpacs,
showkeys,
nofootinbib,
 amsmath,amssymb,
 aps,
altaffilletter,
]{revtex4-1}

\usepackage{graphicx}
\usepackage{multirow}
\usepackage{amsmath}
\usepackage[T1]{fontenc} 
\usepackage{epstopdf} 
\usepackage{bm}
\usepackage{hyperref}
\usepackage[mathlines]{lineno}
\usepackage{xfrac} 
\usepackage{longtable} 
\usepackage{color}
\usepackage{booktabs}
\usepackage{float}
\usepackage{xcolor}
\usepackage[utf8]{inputenc}
\usepackage[english]{babel}
\usepackage{comment}
\usepackage{dcolumn}
\usepackage{siunitx}
\newcolumntype{d}[1]{D{.}{\;\rightarrow\;}{-1}}

\begin{document}
\preprint{APS/123-QED}

\title{High-Precision Branching Ratio Measurement and Spin Assignment Implications for \textsuperscript{62}Ga Superallowed $\ensuremath{\beta}$ Decay}

\author {A.D.~MacLean}
\email{amacle02@uoguelph.ca}
\affiliation{Department of Physics, University of Guelph, Guelph, Ontario N1G 2W1, Canada.}

\author {A.T.~Laffoley}
\email{alaffole@uoguelph.ca}
\address{Department of Physics, University of Guelph, Guelph, Ontario N1G 2W1, Canada.}

\author {C.E.~Svensson}
\affiliation{Department of Physics, University of Guelph, Guelph, Ontario N1G 2W1, Canada.}

\author {G.C.~Ball}
\affiliation{TRIUMF, 4004 Wesbrook Mall, Vancouver, British Columbia V6T 2A3, Canada.}

\author {J.R.~Leslie}
\affiliation{Department of Physics, Queen’s University, Kingston, Ontario K7L 3N6, Canada.}

\author {C.~Andreoiu}
\affiliation{Department of Chemistry, Simon Fraser University, Burnaby, British Columbia V5A 1S6, Canada.}

\author {A.~Babu}
\affiliation{TRIUMF, 4004 Wesbrook Mall, Vancouver, British Columbia V6T 2A3, Canada.}

\author {S.S.~Bhattacharjee}
\affiliation{TRIUMF, 4004 Wesbrook Mall, Vancouver, British Columbia V6T 2A3, Canada.}

\author {H.~Bidaman}
\affiliation{Department of Physics, University of Guelph, Guelph, Ontario N1G 2W1, Canada.}

\author {V.~Bildstein}
\affiliation{Department of Physics, University of Guelph, Guelph, Ontario N1G 2W1, Canada.}

\author {C.~Burbadge}
\affiliation{Department of Physics, University of Guelph, Guelph, Ontario N1G 2W1, Canada.}

\author {M.~Bowry}
\altaffiliation[Present address: ]{Department of Physics, University of the West of Scotland, Glasgow G72 0LH, UK.}
\affiliation{TRIUMF, 4004 Wesbrook Mall, Vancouver, British Columbia V6T 2A3, Canada.}

\author {C.~Cheng}
\affiliation{TRIUMF, 4004 Wesbrook Mall, Vancouver, British Columbia V6T 2A3, Canada.}

\author {D.S.~Cross}
\affiliation{Department of Chemistry, Simon Fraser University, Burnaby, British Columbia V5A 1S6, Canada.}

\author {A.~Diaz-Varela}
\affiliation{Department of Physics, University of Guelph, Guelph, Ontario N1G 2W1, Canada.}

\author {I.~Dillmann}
\affiliation{TRIUMF, 4004 Wesbrook Mall, Vancouver, British Columbia V6T 2A3, Canada.}
\affiliation{Department of Physics and Astronomy, University of Victoria, Victoria, British Columbia V8P 5C2, Canada.}

\author {M.R.~Dunlop}
\affiliation{Department of Physics, University of Guelph, Guelph, Ontario N1G 2W1, Canada.}

\author {R.~Dunlop}
\affiliation{Department of Physics, University of Guelph, Guelph, Ontario N1G 2W1, Canada.}

\author {L.J.~Evitts}
\altaffiliation[Present address: ]{Nuclear Futures Institute, Bangor University, Bangor, Gwynedd, LL57 2DG, UK}
\affiliation{TRIUMF, 4004 Wesbrook Mall, Vancouver, British Columbia V6T 2A3, Canada.}
\affiliation{Department of Physics, University of Surrey, Guildford GU2 7XH, UK}

\author {P.~Finlay}
\altaffiliation[Present address: ]{Xanadu, 777 Bay Street, Toronto, Ontario, M5G 2C8, Canada}
\affiliation{K.U. Leuven, Instituut voor Kern- en Stralingsfysica, Celestijnenlaan 200D, 3001 Leuven, Belgium}

\author {S.~Gillespie}
\affiliation{TRIUMF, 4004 Wesbrook Mall, Vancouver, British Columbia V6T 2A3, Canada.}

\author {A.B.~Garnsworthy}
\affiliation{TRIUMF, 4004 Wesbrook Mall, Vancouver, British Columbia V6T 2A3, Canada.}

\author {P.E.~Garrett}
\affiliation{Department of Physics, University of Guelph, Guelph, Ontario N1G 2W1, Canada.}

\author {E.~Gopaul}
\affiliation{TRIUMF, 4004 Wesbrook Mall, Vancouver, British Columbia V6T 2A3, Canada.}

\author {C.J.~Griffin}
\affiliation{TRIUMF, 4004 Wesbrook Mall, Vancouver, British Columbia V6T 2A3, Canada.}

\author {G.F.~Grinyer}
\affiliation{Department of Physics, University of Regina, Regina, SK S4S 0A2, Canada.}

\author {G.~Hackman}
\affiliation{TRIUMF, 4004 Wesbrook Mall, Vancouver, British Columbia V6T 2A3, Canada.}

\author {J.~Henderson}
\altaffiliation[Present address: ]{Lawrence Livermore National Laboratory, 7000 East Ave, Livermore, California 94550, USA.}
\affiliation{TRIUMF, 4004 Wesbrook Mall, Vancouver, British Columbia V6T 2A3, Canada.}

\author {B.~Jigmeddorj}
\altaffiliation[Present address: ]{Applied Physics Branch, Canadian Nuclear Laboratories, 286 Plant Road, Chalk River, Ontario K0J 1J0, Canada}
\affiliation{Department of Physics, University of Guelph, Guelph, Ontario N1G 2W1, Canada.}

\author {K.G.~Leach}
\affiliation{Department of Physics, Colorado School of Mines, Golden, CO 80401, USA.}

\author {E.~Kassanda}
\affiliation{Department of Physics, University of Guelph, Guelph, Ontario N1G 2W1, Canada.}

\author {J.~McAfee}
\affiliation{TRIUMF, 4004 Wesbrook Mall, Vancouver, British Columbia V6T 2A3, Canada.}
\affiliation{Department of Physics, University of Surrey, Guildford GU2 7XH, UK}

\author {M.~Moukaddam}
\altaffiliation[Present address: ]{Universit\'{e} de Strasbourg, IPHC, 23 rue du Loess, 67037, Strasbourg, France.}
\affiliation{TRIUMF, 4004 Wesbrook Mall, Vancouver, British Columbia V6T 2A3, Canada.}

\author {C.~Natzke}
\affiliation{TRIUMF, 4004 Wesbrook Mall, Vancouver, British Columbia V6T 2A3, Canada.}
\affiliation{Department of Physics, Colorado School of Mines, Golden, CO 80401, USA.}

\author {S.~Nittala}
\affiliation{TRIUMF, 4004 Wesbrook Mall, Vancouver, British Columbia V6T 2A3, Canada.}

\author {B.~Olaizola}
\affiliation{TRIUMF, 4004 Wesbrook Mall, Vancouver, British Columbia V6T 2A3, Canada.}

\author {J.~Park}
\altaffiliation[Present address: ]{Department of Physics, Lund University, 22100 Lund, Sweden.}
\affiliation{TRIUMF, 4004 Wesbrook Mall, Vancouver, British Columbia V6T 2A3, Canada.}

\author {C.~Paxman}
\affiliation{TRIUMF, 4004 Wesbrook Mall, Vancouver, British Columbia V6T 2A3, Canada.}
\affiliation{Department of Physics, Univesity of Surrey, Guildford GU2 7XH, UK}

\author {J.L.~Pore}
\altaffiliation[Present address: ]{Lawrence Berkeley National Laboratory, Berkeley, California 94720, USA.}
\affiliation{Department of Chemistry, Simon Fraser University, Burnaby, British Columbia V5A 1S6, Canada.}

\author {C.~Porzio}
\affiliation{TRIUMF, 4004 Wesbrook Mall, Vancouver, British Columbia V6T 2A3, Canada.}
\affiliation{ L'Istituto Nazionale di Fisica Nucleare (INFN) Sezione of Milano and Dipartimento di Fisica, Universit\`{a} di Milano, Milano, Italy}

\author {A.J.~Radich}
\affiliation{Department of Physics, University of Guelph, Guelph, Ontario N1G 2W1, Canada.}

\author {P.~Ruotsalainen}
\altaffiliation[Present address: ]{University of Jyv{\"a}skyl{\"a}, Department of Physics, P.O. Box 35, Fl-40014, University of Jyv{\"a}skyl{\"a}, Finland}
\affiliation{TRIUMF, 4004 Wesbrook Mall, Vancouver, British Columbia V6T 2A3, Canada.}

\author {Y.~Saito}
\affiliation{TRIUMF, 4004 Wesbrook Mall, Vancouver, British Columbia V6T 2A3, Canada.}
\affiliation{Department of Physics and Astronomy, University of British Columbia, Vancouver, BC, V6T 1Z4, Canada.}

\author {S.~Sharma}
\affiliation{Department of Physics, University of Regina, Regina, SK S4S 0A2, Canada.}

\author {J.~Smallcombe}
\altaffiliation[Present address: ]{Department of Physics, University of Liverpool L69 3BX, Liverpool, UK.}
\affiliation{TRIUMF, 4004 Wesbrook Mall, Vancouver, British Columbia V6T 2A3, Canada.}

\author {J.K.~Smith}
\altaffiliation[Present address: ]{Department of Physics, Pierce College, Puyallup, Washington 98374, USA.}
\affiliation{TRIUMF, 4004 Wesbrook Mall, Vancouver, British Columbia V6T 2A3, Canada.}

\author {R.~Sultana}
\affiliation{TRIUMF, 4004 Wesbrook Mall, Vancouver, British Columbia V6T 2A3, Canada.}

\author {J.~Turko}
\affiliation{Department of Physics, University of Guelph, Guelph, Ontario N1G 2W1, Canada.}

\author {J.~Williams}
\affiliation{TRIUMF, 4004 Wesbrook Mall, Vancouver, British Columbia V6T 2A3, Canada.}

\author {D.~Yates}
\affiliation{TRIUMF, 4004 Wesbrook Mall, Vancouver, British Columbia V6T 2A3, Canada.}
\affiliation{Department of Physics and Astronomy, University of British Columbia, Vancouver, BC, V6T 1Z4, Canada.}

\author {T.~Zidar}
\affiliation{Department of Physics, University of Guelph, Guelph, Ontario N1G 2W1, Canada.}

\date{\today}

\begin{abstract}
A high-precision branching ratio measurement for the superallowed Fermi $\beta^{+}$ emitter $^{62}$Ga was performed with the Gamma-Ray Infrastructure for Fundamental Investigations of Nuclei (GRIFFIN) spectrometer at the Isotope Separator and Accelerator (ISAC) radioactive ion beam facility at TRIUMF. The high efficiency of the GRIFFIN spectrometer allowed 63 $\gamma$-ray transitions, with intensities down to $\approx$1 part per million (ppm) per $^{62}$Ga $\beta^{+}$ decay, to be placed in the level scheme of the daughter nucleus $^{62}$Zn, establishing the superallowed $\beta$ branching ratio for $^{62}$Ga decay to be 99.8577$^{+0.0023}_{-0.0029}\%$, a factor of 4 more precise than the previous world average. For several cascades, $\gamma-\gamma$ angular correlation measurements were performed to assign spins and/or determine the mixing ratios of transitions. In particular, the spin of the 2.342~MeV excited state in the daughter nucleus $^{62}$Zn was definitively assigned as $J = 0$. This assignment resolves a discrepancy between previous measurements and has important implications for the isospin symmetry breaking correction, $\delta_{C1}$, in $^{62}$Ga superallowed Fermi $\beta$ decay. 
\end{abstract}



\maketitle

\section{Introduction}
Superallowed Fermi $\beta$ decays between $J^{\pi} = 0^{+}$ nuclear isobaric analogue states provide stringent tests of the electroweak Standard Model and the most precise determination of the weak vector coupling constant, $G_{V}$~\cite{Hardy2015}. Combined with the Fermi coupling constant, $G_{F}$, from purely leptonic muon decay, the value of $G_{V}$ from superallowed Fermi $\beta$ decays also currently provides the most precise determination of the up-down matrix element, $V_{ud} = G_{V}/G_{F}$, of the Cabibbo-Kobayashi-Maskawa (CKM) quark mixing matrix~\cite{PDG}. To determine $G_{V}$, precise measurements of the $ft$-values for superallowed $\beta$ decays are required, together with theoretical calculations of the small (order 1$\%$) corrections necessary to obtain the transition independent corrected $\mathcal{F}t$-values~\cite{Hardy2015}:
\begin{equation}
\mathcal{F}t = ft(1+\delta_{R}')(1+\delta_{NS}-\delta_{C}) = \frac{K}{2G_{V}^{2}(1+\Delta^{V}_{R})}
\label{eq:Ft}
\end{equation}
where $K/(\hbar c)^{6} = 2\pi^{3}\hbar \ln2/(m_{e}c^{2})^{5} = 8120.27648(26) \times 10^{-10}\text{ GeV}^{-4}\text{s}$, $\delta'_{R}$ and $\delta_{NS}$ are transition-dependent ``outer'' radiative corrections, $\Delta^V_{R}$ is the transition-independent ``inner'' radiative correction, and $\delta_{C}$ is a nucleus-dependent isospin-symmetry-breaking correction. The experimental $ft$ values are determined from precise measurements of the decay $Q$-values, half-lives and branching ratios. 

For more than a decade, surveys of the world superallowed Fermi $\beta$ decay data~\cite{Hardy2015, Towner2010RPP, Hardy2009}, have relied on the value of $\Delta^{V}_{R} = 2.361(38)\%$ calculated in Ref.~\cite{Marciano2006} to extract the value of $V_{ud}$. A recent evaluation of the superallowed data~\cite{Hardy2018} yielded: 
\begin{equation}
\lvert V_{ud}\rvert=0.97420(10)_{\mathcal{F}t}(18)_{\Delta^V_R} = 0.97420(21)
\end{equation}
with an uncertainty dominated by the precision of $\Delta^{V}_{R}$.
Recently, a new calculation~\cite{Seng2018} has significantly reduced the hadronic uncertainty in the determination of $\Delta^{V}_{R}$ by expressing the $\gamma W$-box contribution in terms of a dispersion relation, yielding $\Delta^{V}_{R} = 2.467(22)\%$, with the potential for further reductions in uncertainty through improved lattice calculations~\cite{SengPRL2019}. The larger value of $\Delta^{V}_{R}$ leads to a corresponding reduction in $V_{ud}$:
\begin{equation}
\lvert V_{ud}\rvert=0.97370(10)_{\mathcal{F}t}(10)_{\Delta^V_R} = 0.97370(14).
\end{equation}
Combined with $\lvert V_{us}\rvert = 0.2245(8)$ obtained from a weighted average of the 2020 Particle Data Group~\cite{PDG} values from $K_{\mu 2}$ ($\lvert V_{us} \rvert = 0.2252(5)$) and $K_{l3}$ ($\lvert V_{us} \rvert = 0.2231(7)$) decays with an uncertainty scaled to account for the inconsistency between the two determinations, and $\lvert V_{ub}\rvert = 0.00382(24)$ yields:  
\begin{equation}
\lvert V_{ud}\rvert^{2} + \lvert V_{us}\rvert^{2} + \lvert V_{ub}\rvert^{2} = 0.9985(3)_{V_{ud}}(4)_{V_{us}} = 0.9985(5)
\label{eq:CKM}
\end{equation}
for the sum of the squares of the first row CKM matrix elements. This result violates unitarity at the 3.3$\sigma$ level. 

As a violation of CKM unitarity would require new physics beyond the current electroweak Standard Model ({\em cf.}~Ref.~\cite{Belfatto2020}) all inputs to Eq.~\ref{eq:CKM} must be carefully scrutinized. While a resolution of the inconsistencies in the determination of $V_{us}$ from the $K_{\mu 2}$ and $K_{l3}$ decays would obviously be highly desirable, we note that even adoption of the larger value of $\lvert V_{us}\rvert = 0.2252(5)$ from $K_{\mu 2}$ decays would still leave a 3.3$\sigma$ discrepancy in the unitary sum. For the evaluation of $V_{ud}$, an independent calculation of $\Delta^{V}_{R}$ employing updated perturbative QCD techniques yielded $\Delta^{V}_{R} = 2.426(32)\%$~\cite{CzarneckiPRD2019}, a value intermediate between the original result of Ref.~\cite{Marciano2006} and the recent dispersion relation based calculation of Ref.~\cite{Seng2018}, which would reduce the discrepancy with CKM unitarity to 2.2$\sigma$. Exploratory calculations of the nuclear-structure dependent ``outer'' radiative correction $\delta_{NS}$ in Eq.~\ref{eq:Ft} by dispersion-relation techniques have also recently been performed~\cite{SengPRD2019, Gorchtein2019}. While these calculations suggest that the quenching of the free neutron Born contribution in the nuclear environment has been underestimated in previous calculations, which would move the average $\mathcal{F}t$ value closer to the requirements for CKM unitarity~\cite{SengPRD2019}, this effect is at least partially cancelled by an energy-dependent contribution to $\delta_{NS}$ that has been neglected in previous calculations~\cite{Gorchtein2019}. This leaves the central value of $\overline{\mathcal{F}t}$ effectively unchanged, while increasing its uncertainty and thus reducing the discrepancy with CKM unitarity to 1.9$\sigma$~\cite{Gorchtein2019}. Finally, the isospin symmetry breaking corrections $\delta_{C}$ required to correct the experimental superallowed $ft$-values for the breaking of isospin symmetry by Coulomb and charge-dependent nuclear forces have been the focus of a large number of studies over a period of decades ~\cite{Towner1977, Ormand1985, Ormand1989, Ormand1995, Towner2002, Towner2008, Hardy2009, Miller2008, *Miller2009, Auerbach2009, Liang2009, Grinyer2010, Towner2010, Satula2011, Satula2012, Rodin2013, Xayavong2015, leach2018}. Given the recent advances in the calculation of the radiative corrections for the superallowed Fermi $\beta$ decays and the apparent violation of CKM unitarity in Eq.~\ref{eq:CKM}, continued scrutiny of these model-dependent isospin symmetry breaking corrections, and their role in determining $\mathcal{F}t$ and $V_{ud}$, remains crucial. 

The isospin symmetry breaking correction is typically separated into two components~\cite{Towner2008}: 
\begin{equation}
    \delta_{C} = \delta_{C1} + \delta_{C2}
\end{equation}
where $\delta_{C1}$ accounts for differences in configuration mixing between the parent and daughter 0$^{+}$ states in the superallowed decay and $\delta_{C2}$ accounts for the imperfect radial overlap of the  wavefunctions for the proton in the parent nucleus and the neutron in the daughter nucleus due to differences in binding energy and the Coulomb potential experienced by the protons. The first term, $\delta_{C1}$, is particularly amenable  to experimental verification as it leads to weak $\beta$ branches that populate nonanalogue 0$^{+}$ states in the daughter nucleus.

If the square of the Fermi matrix element to the n\textsuperscript{th} excited 0$^{+}$ state in the daughter nucleus is defined as $2\delta^{n}_{C1}$, the Fermi matrix element squared for the superallowed transition to the analogue 0$^{+}$ state is reduced by a factor ($1-\delta_{C1}$), where we have the approximate relation~\cite{Towner2008}:
\begin{equation}
    \delta_{C1} \approx \sum_{n}\delta^{n}_{C1}.
    \label{eq:eqThree}
\end{equation}
If all 0$^{+}$ states included in the shell-model calculation of $\delta_{C1}$ have the same isospin ($T = 1$ in this case) then the approximation in Eq.~\ref{eq:eqThree} becomes exact. For excited 0$^{+}$ states within the $\beta$ decay $Q$-value window, the $\delta^{n}_{C1}$ values are, in principle, experimentally measurable through~\cite{Hagberg1994} 
\begin{equation}
    \delta_{C1}^{n} \approx \frac{f_{0}}{f_{n}}B_{n}
    \label{eq:eqFour}
\end{equation}
where $B_{n}$ is the $\beta$ branching ratio to the n\textsuperscript{th} excited 0$^{+}$ state and $f_{0}$ and $f_{n}$ are the phase-space integrals for decay to the ground state and n\textsuperscript{th} excited 0$^+$ state, respectively. These experimentally measured $\delta^{n}_{C1}$values can be directly compared with theory to provide stringent tests of the theoretical models of isospin symmetry breaking in superallowed Fermi decays.

In the current work, a high-precision branching ratio measurement for $^{62}$Ga $\beta$ decay was performed to constrain the isospin symmetry breaking corrections for this high-$Z$ superallowed $\beta^+$ emitter. Through $\gamma-\gamma$ angular correlation measurements with the GRIFFIN spectrometer a discrepancy~\cite{Kusakari1972, Hinrichs1974, Fulbright1977, Albers2010, Leach2013, Leach2019} regarding the location of the first excited 0$^{+}$ state in the daughter nucleus $^{62}$Zn was resolved, with important implications for the calculation of $\delta_{C1}$. Additionally, limits on nonanalogue $\beta$ branches to higher-lying excited 0$^{+}$ states were determined and compared with theoretical calculations of the $\delta^{n}_{C1}$ values. 

\section{Experiment}
The $^{62}$Ga radioactive ion beam was produced with a 479-MeV proton beam of 70~$\mu$A intensity delivered by the TRIUMF cyclotron~\cite{BYL14} that impinged on a 11.91~g/cm$^2$ zirconium carbide (ZrC) production target. The TRIUMF Resonant Ionization Laser Ion Source (TRILIS)~\cite{Bricault2014} was used to ionize products created by the spallation reactions, which were then mass separated with $A/q=62$ before being delivered as a low-energy (25 keV) beam to the experimental station at a rate between 6,000 and 12,000 ions/s of $^{62}$Ga. The beam also contained minor isobaric contaminants of surface ionized $^{62}$Cu, $^{62}$Co, and $^{62}$Co\textsuperscript{m}, as described below.

The ions were implanted in a mylar tape system at the center of the GRIFFIN spectrometer~\cite{SVE14,GAR19}. To focus on the decay of short-lived $^{62}$Ga a tape cycling mode was used. The primary cycle consisted of a 1.5~s tape move, 2.5~s of background counting before the beam was implanted, 30~s of counting with the beam on, and finally 4~s counting with the beam blocked, before moving the tape and repeating the cycle. This cycling was designed to maximize the statistics for the decay of $^{62}$Ga, with a half-life of $116.121(21)$~ms~\cite{Grinyer2008}, while suppressing the activity of the longer lived daughter, $^{62}$Zn ($T_{1/2}=9.193(15)$~h~\cite{NIC12}), and beam contaminants of $^{62}$Cu ($T_{1/2}=9.67(3)$~min~\cite{NIC12}), $^{62}$Co ($T_{1/2}=1.54(10)$~min~\cite{NIC12}), and $^{62}$Co\textsuperscript{m} ($T_{1/2}=13.86(9)$~min~\cite{NIC12}) observed in the $\gamma$-ray data. Once a cycle was completed, the tape was moved outside of the array and behind a lead shielding wall to prevent contamination of the spectra from these long-lived activities. 
Data were collected in list mode for all $\gamma$-ray and $\beta$-particle events. Gamma-gamma, $\gamma-\beta$ and $\gamma-\gamma-\beta$ coincidences were constructed with timing gates set in the analysis software offline after the data were collected.
\begin{figure}[ht!]
\begin{center}
\includegraphics[width=\columnwidth, keepaspectratio]{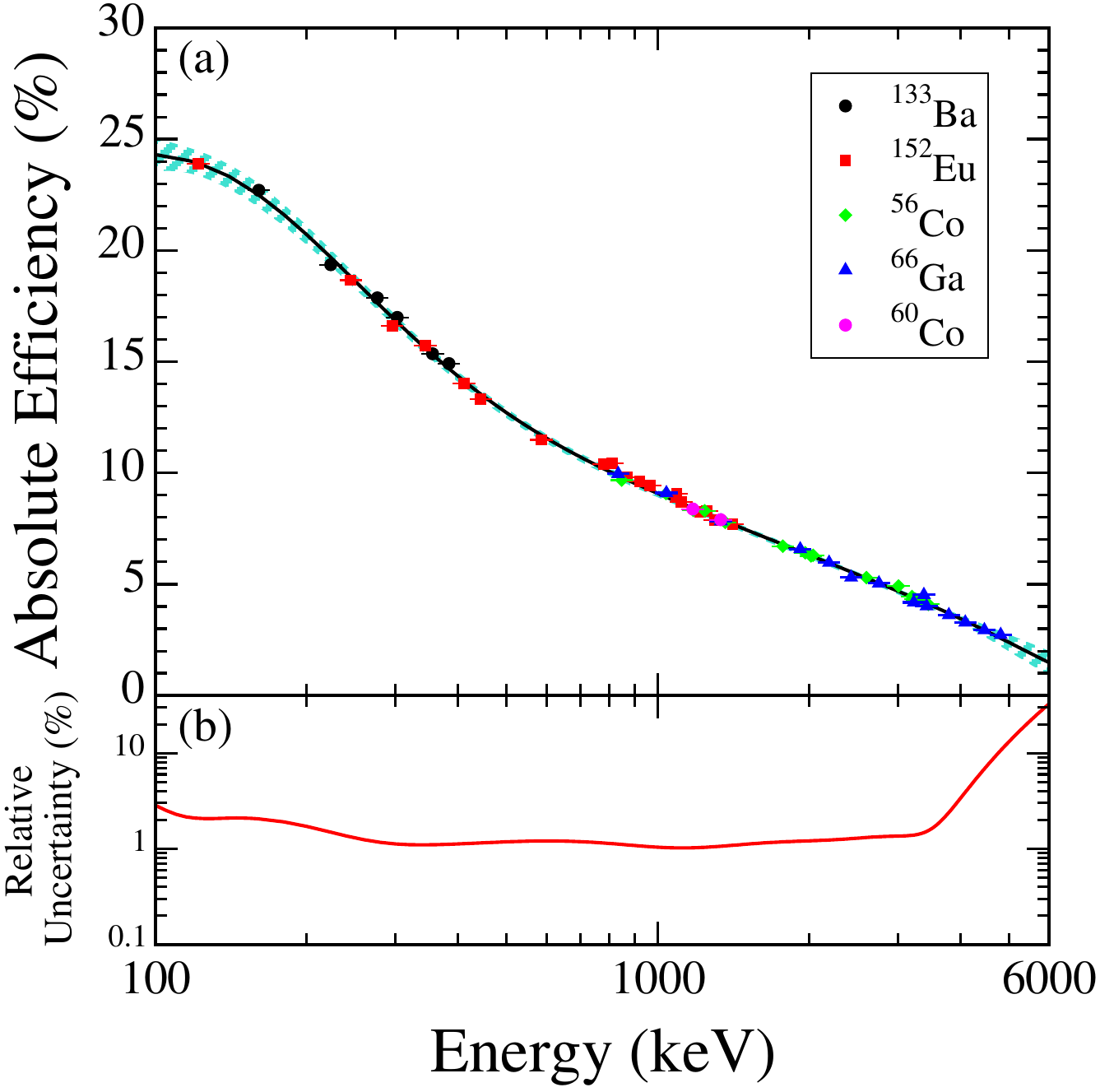}
\caption{(Colour online) (a) Absolute $\gamma$-ray photopeak efficiency of the GRIFFIN spectrometer operated in ``single crystal'' mode obtained using $^{56}$Co, $^{60}$Co, $^{133}$Ba and $^{152}$Eu calibration sources as well as a radioactive ion beam of $^{66}$Ga. The black line is the fit to the data points using methods described in Ref.~\cite{Raman2000} and the turquoise-hatched band represent the uncertainty in the absolute efficiency. (b) Relative  uncertainty in the absolute efficiency calibration.}
\label{fig:efficiency}
\end{center}
\end{figure}

GRIFFIN~\cite{GAR19, SVE14} is an array of 16 high-purity germanium (HPGe) clover detectors~\cite{RIZ16} arranged in a rhombicubocatahedral geometry. Each of the 16 detectors contains four HPGe crystals arranged in a four-leaf clover geometry within a common cryostat. Each HPGe clover is surrounded by a 20 element Compton suppression shield made of bismuth germanate (BGO). For the detection of $\beta$ particles, the SCintillating Electron-Positron Tagging ARray, SCEPTAR,~\cite{GAR19} was used. SCEPTAR is an array of 20 thin (1.6~mm) plastic scintillators, located inside the vacuum chamber surrounding the tape at the center of GRIFFIN. With a $\beta$ detection efficiency of approximately 80$\%$, SCEPTAR is used to provide $\beta$ coincidences with $\gamma$-rays to suppress the effect of signals in GRIFFIN created by room background $\gamma$-ray radiation. The GRIFFIN vacuum chamber was surrounded by a 20-mm thick shell of Delrin plastic to stop the high-energy $\beta$ particles from $^{62}$Ga decay from reaching the HPGe detectors. The low-$Z$ stopping material also limits the bremsstrahlung radiation incident on the GRIFFIN HPGe detectors. The signals from all detectors were continuously digitized by the GRIFFIN digital data acquisition system~\cite{GAR17} with a 100~MHz sampling frequency. A digital implementation of a constant-fraction discriminator algorithm gave timestamps with a precision of $\approx$1~ns following interpolation. To determine the relative efficiency calibration of the GRIFFIN spectrometer as a function of $\gamma$-ray energy, standard sources of $^{152}$Eu, $^{133}$Ba, and $^{56}$Co were used immediately before and following the experiment. As the decay of $^{62}$Ga produces many high energy $\gamma$-rays, a radioactive ion beam of $^{66}$Ga (T$_{1/2}$ = 9.49(3)~hr~\cite{BRO10}) was also implanted in the array at the end of the experiment and counted over several days to extend the relative efficiency calibration to higher energies~\cite{Raman2000}. This relative efficiency calibration was converted to the absolute efficiency curve shown in Fig.~\ref{fig:efficiency} by normalizing to two calibrated $^{60}$Co sources, each with an activity known to $\pm$3\% at a 99\% confidence level. 

\section{Results}
\label{sec:results}

\subsection{Beta Analysis}
\begin{figure}[ht!]
\begin{center}
\includegraphics[width=\columnwidth, keepaspectratio]{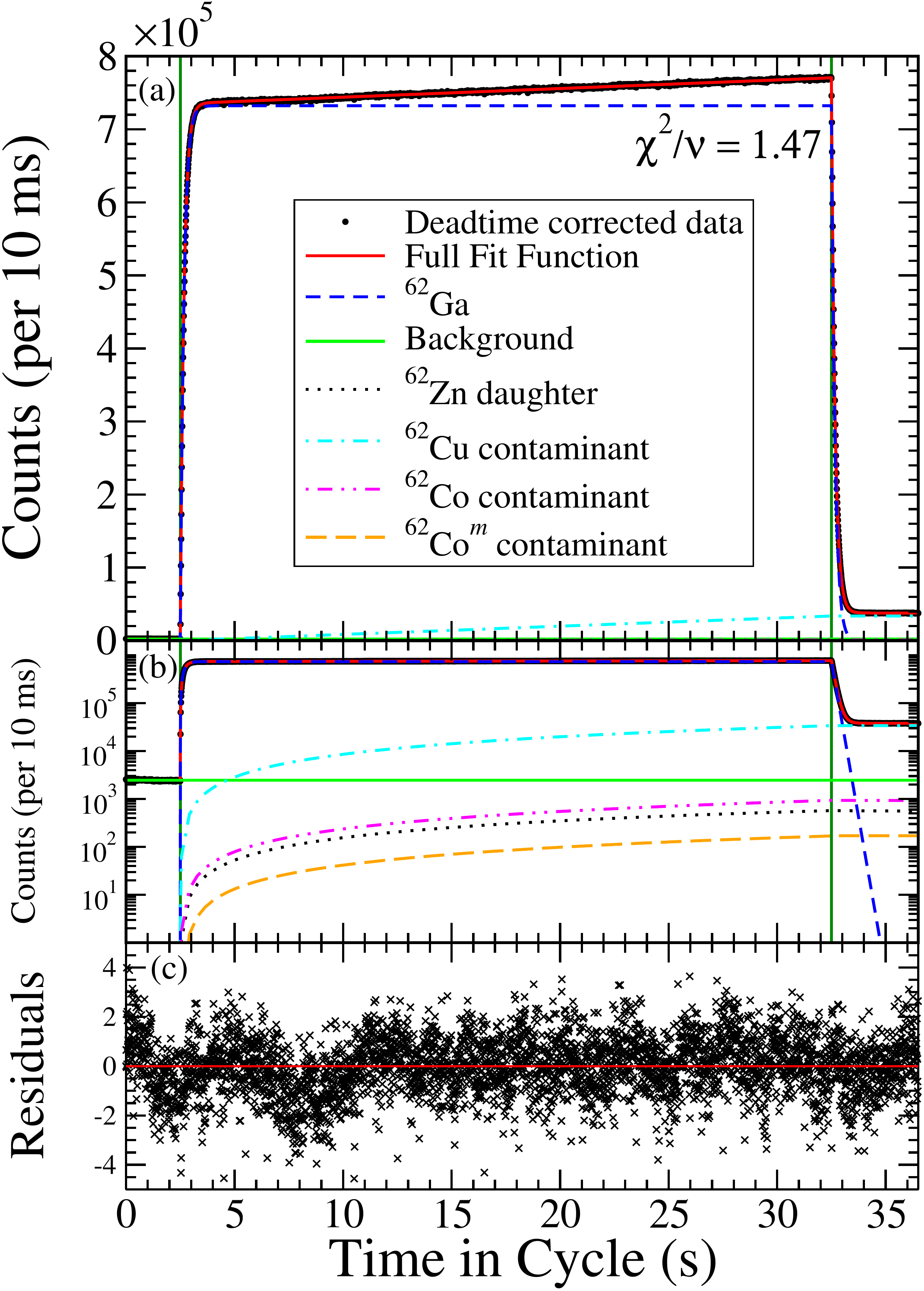}
\caption{(Colour online) Summed $\beta$ activity measured with SCEPTAR over the 8560 cycles used in the analysis on a linear (a) and logarithmic (b) scale. The first 2.5 seconds of the cycle involved background counting with no beam. From 2.5 seconds to 32.5 seconds the beam was turned on, as indicated by the vertical green lines. In the last 4 seconds the beam was turned off and the implanted radioactive beam decays. The tape was moved over a period of 1.5~s between each cycle (not shown). Panel (a) highlights the fit to the entire $\beta$ activity curve, while panel (b) illustrates the very small contributions of the contaminants to the total $\beta$ activity. Panel (c) shows the residuals to the fit, with a line indicating 0.}
\label{fig:BetaCycles}
\end{center}
\end{figure}

The analysis of the $\beta$ activity as a function of time within the cycle was used to determine the total number of decays of $^{62}$Ga observed in the current experiment in order to obtain absolute $\gamma$-ray intensities per $^{62}$Ga $\beta$ decay. The data were first analyzed on a cycle-by-cycle basis and cycles during which the beam intensity was not constant (due to a loss of protons on the production target) were removed. Additionally, a pulse height threshold was applied offline and was used for all $\beta$ singles and $\beta-\gamma$ coincidences to remove low-energy noise events in the $\beta$ detectors. 
In total, 8560 cycles comprising 95$\%$ of the total events recorded  were used in the final analysis with an average beam rate of 10,400 $^{62}$Ga ions/s being delivered to the array during these cycles.  A fixed dead time of 2~$\mu$s following the detection of a $\beta$ particle was applied to the entire SCEPTAR array in the offline analysis to ensure that $\beta$ particles that scattered between SCEPTAR detectors were not multiply counted. By imposing a fixed dead time per event, the observed data could be corrected to obtain the true number of $\beta$-particle events following the techniques described in Refs.~\cite{KOSLOWSKY1997,Grinyer2005}. The dead-time correction was a maximum of 1.8\% at the peak of the $\beta$-activity curve. Following cycle selection and dead-time correction, the data were summed over the 8560 accepted cycles and the summed $\beta$-activity curve shown in Fig.~\ref{fig:BetaCycles} was fit with a maximum-likelihood procedure based on a Leven-Marquardt algorithm~\cite{Grinyer2005}.

The fit function used the fixed beam on and beam off times and included components for the $^{62}$Ga beam and the grow-in of its daughter $^{62}$Zn, as well as beam contaminants of $^{62}$Cu, $^{62}$Co, and $^{62}$Co\textsuperscript{m} and a constant room background. The background rate was fixed to the value observed during the ``background'' phase of the cycle, after the tape was moved (to remove the long-lived contaminants) but before the beam was delivered and was observed to be $\approx$30 counts/s summed over the 20 SCEPTAR detectors. The half-lives for each of the decaying components were fixed to their literature values from Ref.~\cite{NIC12}. The relative intensities of the long-lived in-beam contaminants were determined from a $\gamma$-ray analysis to be in the ratio $^{62}$Co:$^{62}$Co\textsuperscript{m}:$^{62}$Cu = 1.0(3):1.00(2):74(6) and were fixed, leaving only two free parameters: the intensity of the $^{62}$Ga beam and the summed intensity of the contaminants.
The fit to the $\beta$-activity curve shown in Fig.~\ref{fig:BetaCycles} yielded a total of $N_{\beta} = 2.1960(14)\times 10^9$ $^{62}$Ga $\beta$-decays detected during the experiment, where the uncertainty includes a contribution from varying the fixed half-lives for $^{62}$Ga, $^{62}$Cu, $^{62}$Co, and $^{62}$Co\textsuperscript{m} within their $\pm1\sigma$ uncertainty ranges. A freeing of the relative intensities of the beam contaminants was also performed as a systematic check and had a negligible effect on the $\gamma$ intensities and $\beta$ branches reported in Section~\ref{sec:gamma}.

\begin{figure*}[ht!]
\begin{center}
\includegraphics[width=0.85\textwidth, keepaspectratio]{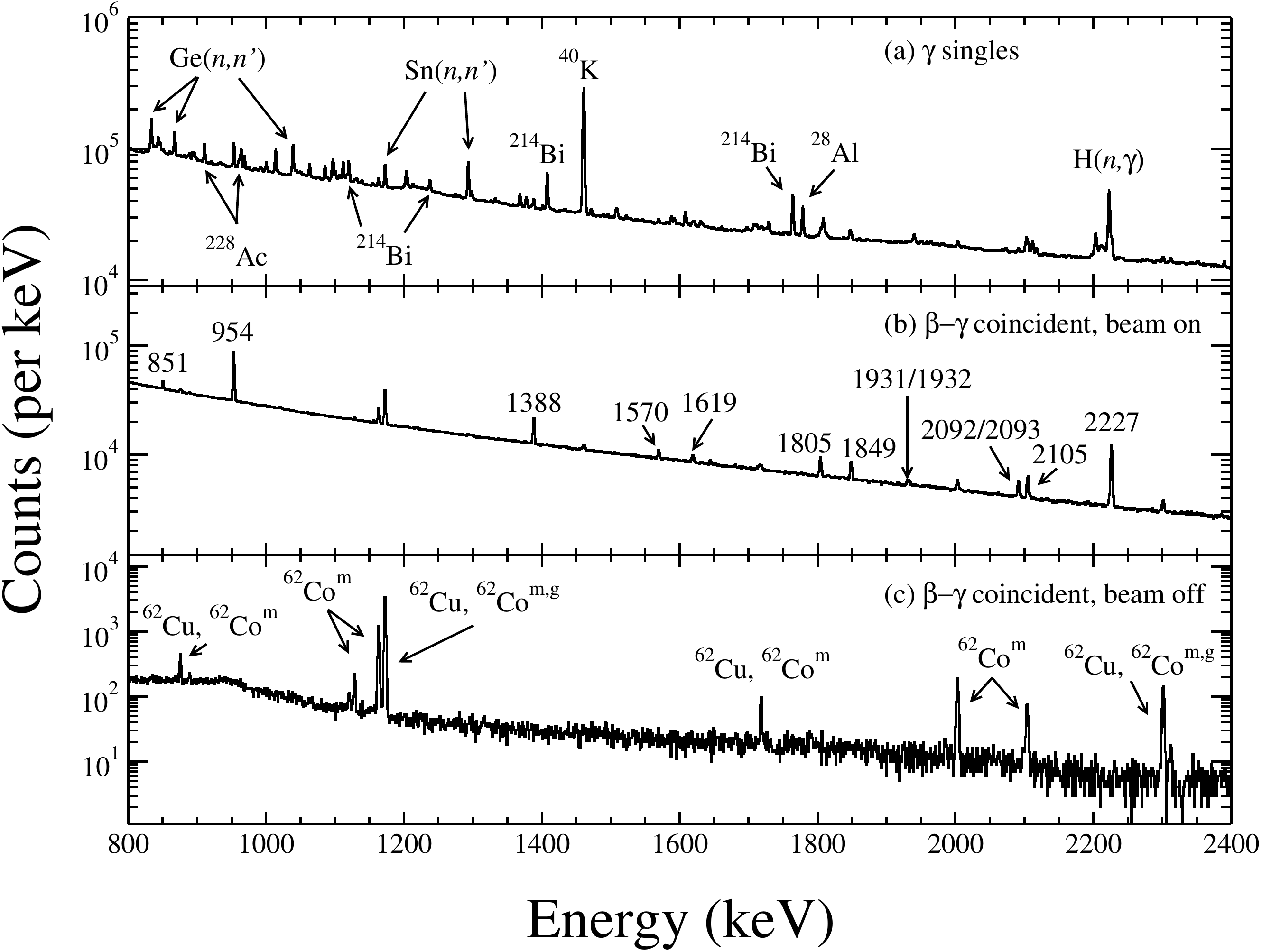}
\caption{Gamma-ray spectra of (a) $\gamma$-singles without a $\beta$ coincidence, (b) $\beta-\gamma$ coincident events during the beam-on time, and (c) $\beta-\gamma$ coincident events during the last 2.5~seconds of the cycle, after the $^{62}$Ga activity had decayed to a negligible level.}
\label{fig:beamOn}
\end{center}
\end{figure*}

\begin{figure*}[ht!]
\begin{center}
\includegraphics[width=0.80\textwidth, keepaspectratio]{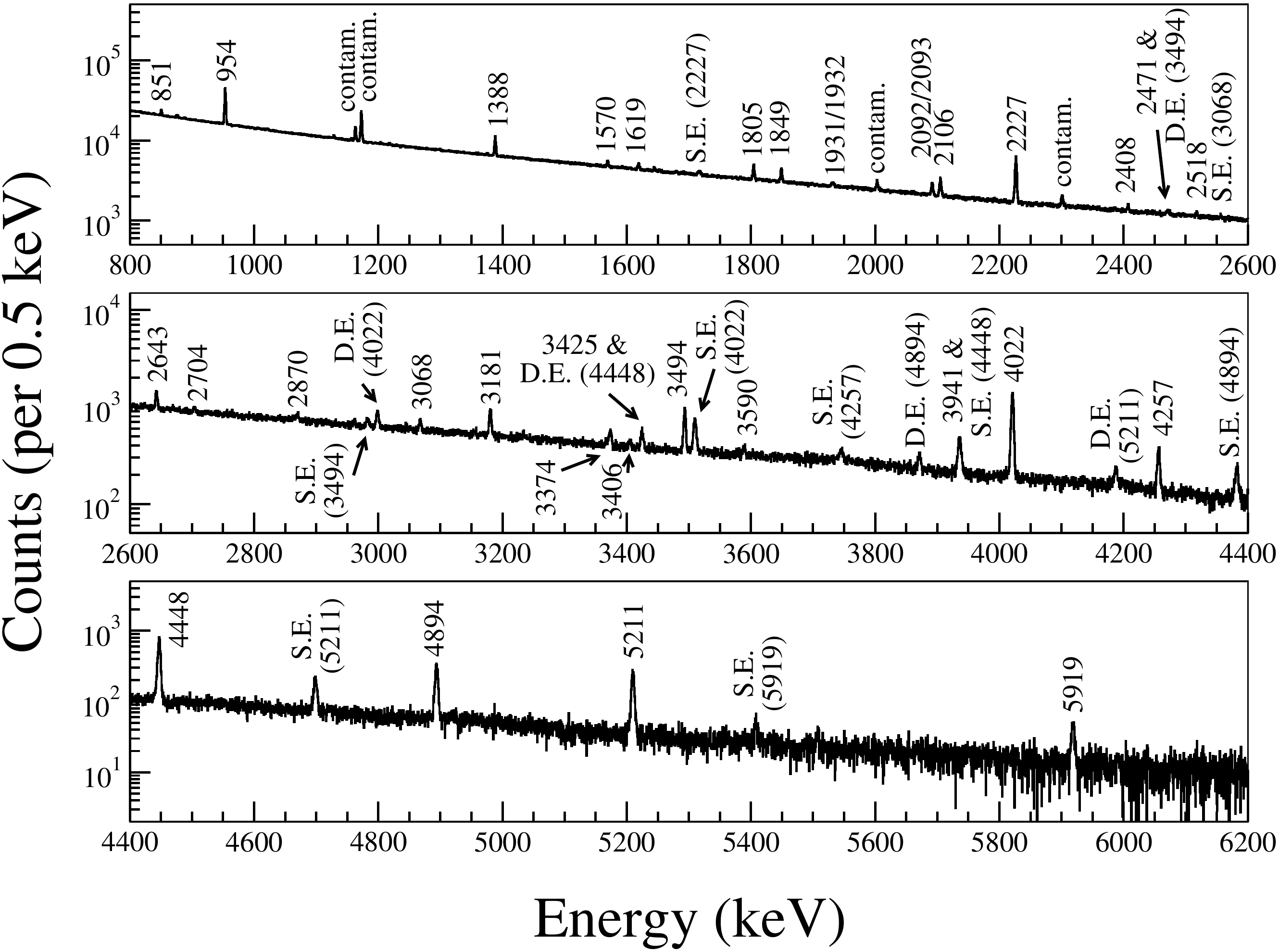}
\caption{The spectrum of $\gamma$-rays in coincidence with a $\beta$ particle detected in SCEPTAR during the beam-on portion of the cycle. Single- and double-escape peaks are labeled as S.E. and D.E., respectively, and the full energy is indicated.}
\label{fig:singles}
\end{center}
\end{figure*}

\subsection{Gamma-ray Intensities}
\label{sec:gamma}
The $\gamma$-ray data were analyzed in $\beta$-tagged $\gamma$ singles as well as $\beta$-tagged $\gamma-\gamma$ coincidences. Figure~\ref{fig:beamOn}(a) shows the $\gamma$-ray singles spectrum without a $\beta$ coincidence which, due to the dominant superallowed decay of $^{62}$Ga to the ground state of $^{62}$Zn, is dominated by room background $\gamma$-rays. Figure~\ref{fig:beamOn}(b) shows the corresponding spectrum in coincidence with $\beta$ particles detected in SCEPTAR and is now dominated by $\gamma$-ray transitions following weak Gamow-Teller and nonanalogue Fermi $\beta$-decay branches of $^{62}$Ga to excited states in $^{62}$Zn. Gamma rays from the in-beam contaminants $^{62}$Cu, $^{62}$Co, and $^{62}$Co\textsuperscript{m} are clearly identified in Fig.~\ref{fig:beamOn}(c) gated on the final 2.5~seconds of the cycle after the activity of the short-lived $^{62}$Ga component had decayed to a negligible level. From this analysis, a total of 63 $\gamma$-rays were identified following $^{62}$Ga $\beta$ decay to 24 excited states in the daughter nucleus $^{62}$Zn. This represents an approximate doubling compared to the previous work of Ref.~\cite{Finlay2008}, in which 30 observed $\gamma$-rays were assigned to 13 excited states in $^{62}$Zn. 

\begin{figure*}[ht!]
\begin{center}
\includegraphics[width=0.85\textwidth, keepaspectratio]{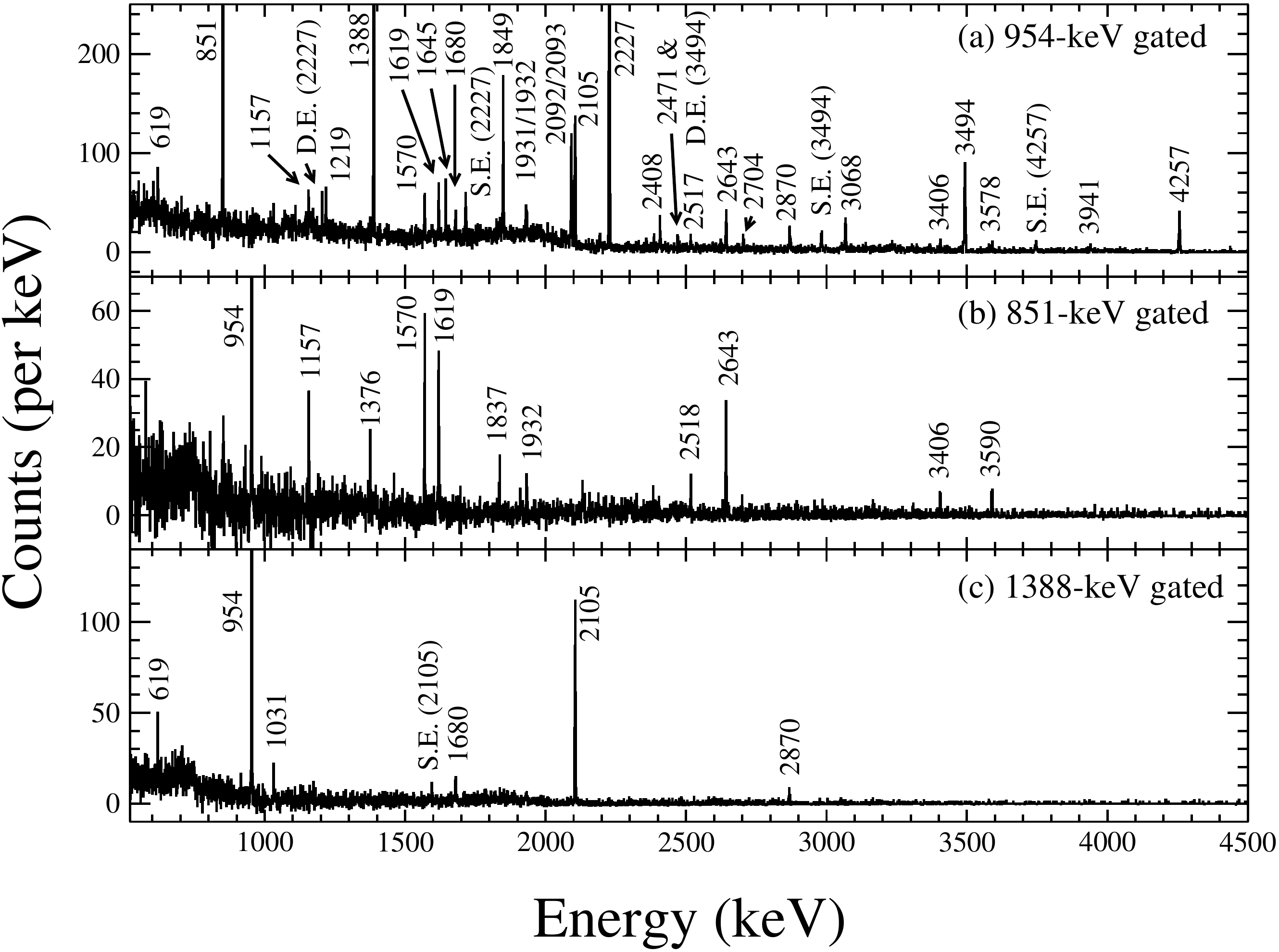}
\caption{Spectra of background subtracted $\beta$-tagged $\gamma$-rays in coincidence with (a) the 954-keV $\gamma$-ray, (b) the 851-keV $\gamma$-ray, and (c) the 1388-keV $\gamma$-ray. Single- and double-escape peaks are labeled as S.E. and D.E., respectively, and the full energy is indicated.}
\label{fig:coinc954}
\end{center}
\end{figure*}

\begin{figure*}[ht!]
\begin{center}
\includegraphics[width=\textwidth,keepaspectratio]{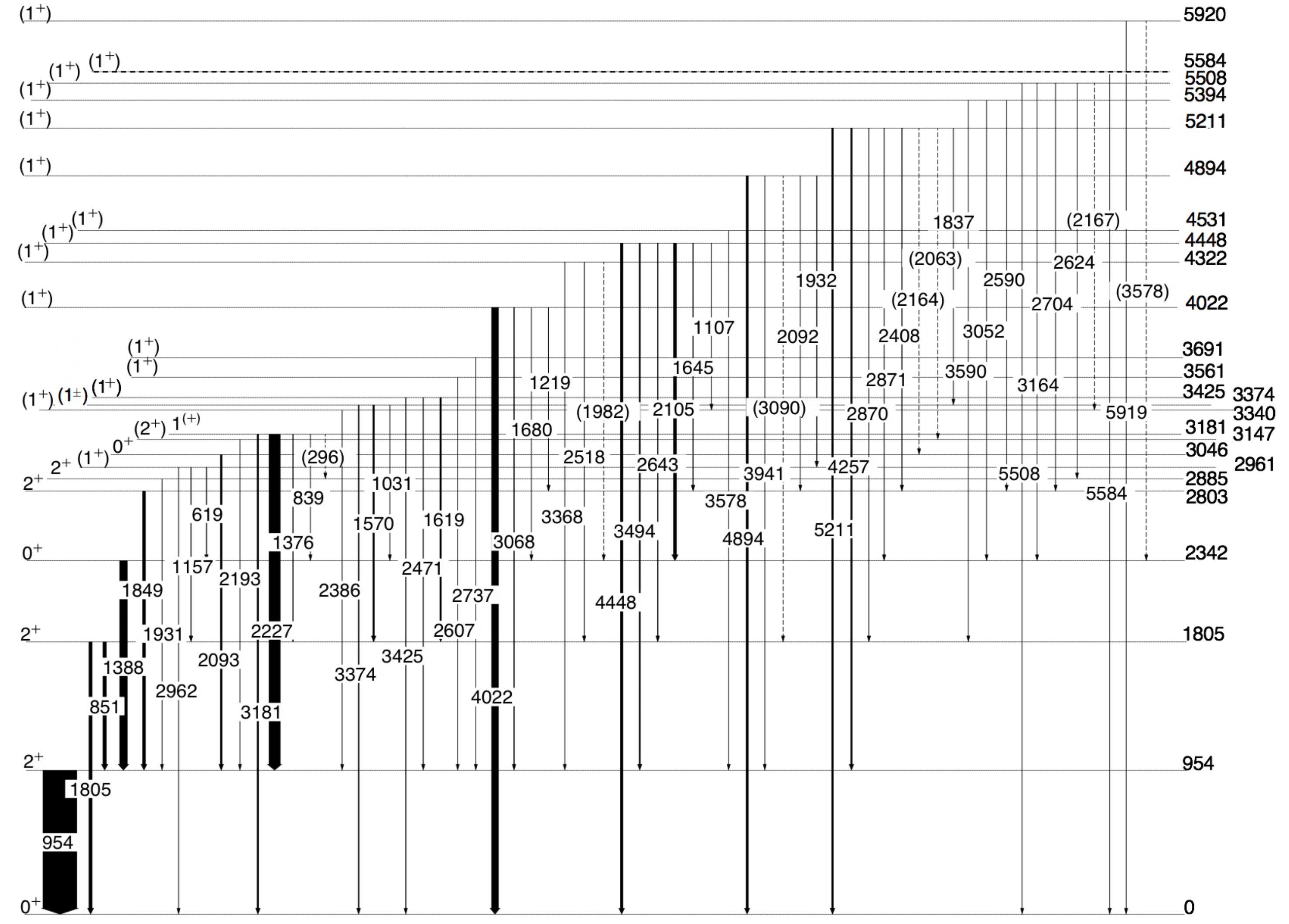}
\caption{Level scheme of $^{62}$Zn populated in the $\beta^{+}$ decay of $^{62}$Ga. The widths of the arrows indicate the relative intensities of the transitions. Transitions with intensities that are consistent with zero at the 1.5$\sigma$ level are shown as tentative with dashed arrows. Additionally, given that the only observed transition from the proposed 5584-keV level was directly to the ground state, this level is also labelled as tentative (dashed).}
\label{fig:LevelScheme}
\end{center}
\end{figure*}

If the depopulating $\gamma$-rays were sufficiently intense, their intensities were determined from the $\beta$-tagged $\gamma$-ray singles spectra shown in Figure~\ref{fig:singles}. Summing corrections were applied to the intensities of all transitions following the methods described in Ref.~\cite{GAR19}. 
The effect of the $\beta$ energy threshold applied to SCEPTAR was examined in detail in the work of Finlay \textit{et al.}~\cite{Finlay2008}. This threshold introduces a small bias in the $\gamma$-ray intensity as a function of the $Q_{\beta}$ value for the $\beta$ branch of interest. From the extensive Geant4 simulations performed in Ref.~\cite{Finlay2008}, we determine this systematic effect to be approximately 1\% for a level at 6~MeV excitation energy. At this excitation energy, the data reported in the current work (Table~\ref{tab:gione}) have uncertainties associated with counting statistics and efficiency calibrations that are greater than 10\%, making the $\beta$ threshold bias entirely negligible. As the level excitation energy decreases, the $Q_{\beta}$ value increases and the $\beta$ threshold bias rapidly decreases. For states below 3~MeV, this $\lesssim$0.1\% effect is more than an order of magnitude smaller than the uncertainties associated with counting statistics and the $\gamma$-ray efficiency calibration. Thus, effects associated with the $\beta$ energy threshold are considered  negligible in the current work.
Data were rejected for pile-up, and all intensities reported have been corrected for this effect~\cite{GAR19}. A 7~$\mu$s integration window was used to evaluate peaks, where the data were flagged offline for pile-up if successive hits within the same crystal occurred between the deadtime window, occurring in the first 1.2~$\mu$s after the first hit, and the 8~$\mu$s differentiation time window. Given the beam cycle structure (see Fig.~\ref{fig:BetaCycles}) the pile-up probability was averaged across the beam on time and applied to the data, corresponding to a 0.9\% correction.
In addition, to help reduce the $\gamma$-ray background, a bremsstrahlung-suppression algorithm was applied offline. For each $\beta$ particle detected in the SCEPTAR array, all coincident $\gamma$-rays detected in the GRIFFIN crystals behind that paddle were rejected. This corresponded to an average rejection of approximately 8.4 GRIFFIN crystals per event, and the absolute $\gamma$-ray photopeak efficiency of the spectrometer (Fig.~\ref{fig:efficiency}) was corrected accordingly. 
For weakly populated transitions that could not be observed without performing a $\beta-\gamma-\gamma$ coincidence analysis, the branching ratios were determined by gating on stronger transitions from lower-lying levels to determine their intensities. Examples of background subtracted coincidence $\gamma$-ray spectra are shown in Fig.~\ref{fig:coinc954} and can be compared with Fig.~6 of Ref.~\cite{Finlay2008} to illustrate the large gain in $\gamma-\gamma$ coincidence efficiency provided by the GRIFFIN spectrometer compared to the earlier work performed by our collaboration. The level scheme of $^{62}$Zn populated in the $\beta$ decay of $^{62}$Ga, as deduced in the current work, is shown in Fig.~\ref{fig:LevelScheme}. For this work, we use the convention of identifying the ground state as $0^{+}_{gs}$ while the n\textsuperscript{th} excited $0^{+}$ state is expressed as $0^{+}_{n}$. 

All of the $\gamma$-rays reported in Ref.~\cite{Finlay2008} were confirmed with the exception of the 2802-keV transition, which was reported with an intensity of 6(4)~ppm in Ref.~\cite{Finlay2008} but was not observed in the current work. Two transitions were, however, misplaced in the previous study~\cite{Finlay2008}. These are the 1619-keV and 1157-keV transitions, which were reported in Ref.~\cite{Finlay2008} to originate from a level at 3961~keV. In the current analysis the locations of these transitions are confidently assigned through $\gamma-\gamma$ coincidences. Contrary to the placement proposed in Ref.~\cite{Finlay2008}, the 1619-keV and 1388-keV transitions are not in coincidence with each other, as shown in Fig.~\ref{fig:coinc954}(c), but the 1619-keV $\gamma$ ray is rather in coincidence with the 851-keV $\gamma$ ray as shown in Fig.~\ref{fig:coinc954}(b). Similarly, for the 1157-keV transition there is no observation of a coincidence with the 1849-keV $\gamma$ ray but there is a coincidence with the 851-keV transition as shown in Fig.~\ref{fig:coinc954}(b). These coincidences allowed the 1157-keV and 1619-keV transitions to be confidently assigned as decays from new levels at excitation energies of 2961~keV and 3425~keV respectively, which were corroborated by additional transitions observed decaying from these levels, as shown in Fig.~\ref{fig:LevelScheme}.

An escape peak analysis was performed to ensure that single- and double-escape peaks were not obscuring actual transitions within the level scheme. This was done by measuring the ratio between the full-energy and single-escape peak areas (and those of the full-energy and double-escape peaks) for all $\gamma$~rays of sufficient intensity above 2~MeV. From this analysis, the expected smooth trend with energy was observed with the exception of two data points for the 2471-keV and 3425-keV transitions (corresponding to the double-escape peaks from the intense 3494-keV and 4448-keV transitions). The excess counts indicated that the 2471-keV and 3425-keV peaks also contained actual photopeaks representing the decays of the 3425-keV level to the 954-keV excited state and the ground state, respectively. The smooth trend in the escape-peak relative intensities, along with the intensities of the full-energy peaks were used to remove the double-escape components from these peaks and determine the intensities of the 2471-keV and 3425-keV transitions. In addition, the placement of the new 3425-keV level was further supported by the observation of a feeding transition of 1838~keV from the previously known 5211-keV level.

For the newly identified level at 2961~keV excitation energy, a direct transition to the ground state was observed in addition to 619-keV and 1157-keV transitions to the 0$^{+}_{1}$ and 2$^{+}_{2}$ levels. A feeding transition of 1932~keV from the 4894-keV excited state to the 2961-keV excited state was also observed, giving high confidence in the placement of this new excited state in the level scheme of $^{62}$Zn. The measured $\gamma$-ray intensity balance, $I_{out} - I_{in}$, for this state (see Table~\ref{tab:gione}) was 14(4)~ppm suggesting direct feeding from the $\beta$ decay of $^{62}$Ga. These observations, combined with the fact that the 2961-keV level was not observed in detailed $^{64}$Zn($p,t$)$^{62}$Zn transfer reactions~\cite{Leach2013, Kusakari1972, Hinrichs1974, Leach2019, Fulbright1977} lead to a tentative assignment of ($1^{+}$) for this newly identified state in $^{62}$Zn. 


The 3425-keV state decays to the $0^{+}$ ground state and both the $2^{+}_{2}$ and $2^{+}_{1}$ excited states and has a significant feeding of 64(5)~ppm directly from the $\beta$ decay of $^{62}$Ga and, similar to the state at 2961~keV, was never previously observed in transfer reaction experiments, suggesting a tentative spin assignment of $(1^{+})$ for this state. 

A doublet of $\gamma$-rays was identified at 1931 and 1932~keV, corresponding to transitions from the 2$^{+}_{4}$ state at 2885~keV and the higher lying (1$^{+}$) state at 4894~keV, respectively. Gating on the 1931-keV side of the peak, transitions at 296~keV and 2624~keV were observed feeding the 2884-keV level from excited (1$^{+}$) states, further solidifying this as the previously identified $2_4^+$ level in $^{62}$Zn~\cite{Albers2010}. Another doublet was also observed for the 2105-keV transition, with one component of the transition from the 4448-keV excited level in $^{62}$Zn following the $\beta$ decay of $^{62}$Ga and the other component from $^{62}$Ni following the $\beta$ decay of the contaminant $^{62}$Co\textsuperscript{m}. Using the known intensities of 18.2(6)\% for the 2003.7-keV and 6.7(6)\% for the 2104.9-keV $\gamma$-rays in $^{62}$Ni~\cite{NIC12}, the intensity of the 2105-keV transition in $^{62}$Zn following $^{62}$Ga $\beta$ decay was corrected for the contribution of the $^{62}$Ni contaminant.

An additional doublet involving $\gamma$-rays of 2089~keV and 2092~keV was previously reported in Ref.~\cite{Finlay2008}. The doublet nature of this peak was confirmed in the present work although the $\gamma$-ray energies were revised to 2092~keV and 2093~keV and important corrections were made to the two intensities. In the work of Ref.~\cite{Finlay2008}, a 3042-keV state (3046~keV in the current work) was proposed to decay to the 2$^{+}_{1}$ state via the 2089-keV transition with an intensity of 12(5)~ppm, while the intensity of the 2093-keV transition arising from the 4894-keV state was assigned an intensity of 47(6)~ppm. In the current work, a $\gamma$-ray coincidence gate set on the 1849-keV transition allowed the isolation of the 2092-keV transition arising from the 4894-keV level and a clean determination of its intensity. Fixing this intensity and energy, a fit of the 2092/2093-keV doublet with two peaks established the energy and intensity of the 2093-keV component. Of particular note is that the relative intensities determined in the current work are nearly exchanged compared to those proposed in Ref.~\cite{Finlay2008}, while the sum of their intensities, 52(2)~ppm in this work compared to 59(8)~ppm in Ref.~\cite{Finlay2008}, are in good agreement. The revision of the intensity of the 2093-keV transition decaying from the 0$^{+}_{2}$ excited state at 3046~keV from 12(5)~ppm in Ref.~\cite{Finlay2008} to 46(2)~ppm in the current work implies a higher direct $\beta$ feeding to the 0$^+$ level at 3046~keV and has important consequences for limits on the isospin-symmetry breaking correction $\delta_{C1}^{2}$, as will be discussed in Section~\ref{sec:isospin} below. 

The 3374-keV excited state was not populated strongly enough in the current experiment to allow a $\gamma-\gamma$ angular measurement to be performed to assign the spin of the state. The spin and parity of this state was assigned as 1$^{-}$ in Ref.~\cite{Leach2019}. Additionally, a state at 3.35(2)~MeV was observed in Ref.~\cite{Hinrichs1974} which was tentatively assigned as (1$^-$). In the current work, the measured I$_{out}$ - I$_{in}$ for this state of 61(3)~ppm yields $\log(ft) = 6.62(2)$ assuming negligible unobserved $\gamma$-ray feeding from above. This result does not allow for a definitive assignment of the parity of this state between a 1$^{+}$ state fed by an allowed Gamow Teller decay and a 1$^{-}$ state fed by a first forbidden decay~\cite{Singh1998}. In the current work we thus conservatively adopt $J^\pi = (1^\pm)$ for the spin-parity of the 3374-keV level.

The $\gamma$-ray intensities per $\beta$ decay of $^{62}$Ga deduced in the current work are listed in Table~\ref{tab:gione} and are compared with the results of Ref.~\cite{Finlay2008}.

\begin{center}
\renewcommand{\arraystretch}{0.6}
\begin{longtable*}{S[table-format = 5.4] S[table-format = 5.5]rcS[table-format = 2.4] S[table-format = 2.2] S[table-format = 2.4]}
\caption{Gamma-ray intensities, $I_\gamma$, per $^{62}$Ga $\beta$ decay. E$_{i}$ is the initial level in the transition, E$_{\gamma}$ is the $\gamma$-ray energy, and E$_{f}$ is the final level in the transition. If the sixth column is empty, the transition was not observed in the previous work of Ref.~\cite{Finlay2008}. The final column gives the difference in the observed $\gamma$ decay and feeding intensities for each level in the current work and represents the sum of the direct $\beta$ feeding and unobserved $\gamma$-ray feeding to the level.}
\label{tab:gione} \\

\hline \hline
    \multicolumn{1}{c}{E$_{i}$} & \multicolumn{1}{c}{E$_{\gamma}$}  & \multicolumn{1}{c}{E$_{f}$} & \multicolumn{1}{D{.}{\;\rightarrow\;}{-1}}{J^{\pi}_{i}.J^{\pi}_{f}} &  \multicolumn{1}{c}{$I_{\gamma}$ (ppm)} & \multicolumn{1}{c}{$I_{\gamma}$ (ppm)} & \multicolumn{1}{c}{$I_{out} - I_{in}$}  \\
    \multicolumn{1}{c}{(keV)} & \multicolumn{1}{c}{(keV)} & \multicolumn{1}{c}{(keV)} &  & \multicolumn{1}{c}{This work} & \multicolumn{1}{c}{Ref.~\cite{Finlay2008}} & \multicolumn{1}{c}{(ppm)} \\
\hline \hline                                  
\endfirsthead

\multicolumn{7}{c}{{\bfseries \tablename\ \thetable{} -- continued from previous page}} \\
\hline \hline
    \multicolumn{1}{c}{E$_{i}$} & \multicolumn{1}{c}{E$_{\gamma}$}  & \multicolumn{1}{c}{E$_{f}$} & \multicolumn{1}{D{.}{\;\rightarrow\;}{-1}}{J^{\pi}_{i}.J^{\pi}_{f}} &  \multicolumn{1}{c}{$I_{\gamma}$ (ppm)} & \multicolumn{1}{c}{$I_{\gamma}$ (ppm)} & \multicolumn{1}{c}{$I_{out} - I_{in}$}  \\
    \multicolumn{1}{c}{(keV)} & \multicolumn{1}{c}{(keV)} & \multicolumn{1}{c}{(keV)} & & \multicolumn{1}{c}{This work} & \multicolumn{1}{c}{Ref.~\cite{Finlay2008}} & \multicolumn{1}{c}{(ppm)} \\
\hline \hline  
\endhead

\hline
\multicolumn{7}{c}{Continued on next page} \\
\hline
\endfoot

\hline 
\hline
\multicolumn{5}{r}{\footnotesize $^1$ See text for a discussion of the intensities of the 2092/2093~keV doublet.}
\endlastfoot

       953.7(1) & 953.7(1) & 0 & \multicolumn{1}{D{.}{\;\rightarrow\;}{-1}}{2^+_1.0^+_{gs}} & 839(14)  & 850(19) & 0(16) \\
       1804.5(1) & 850.7(1) & 953.7 & \multicolumn{1}{D{.}{\;\rightarrow\;}{-1}}{2^+_2.2^+_1} & 88(3) & 100(7) & -2(7)  \\
        & 1804.6(1) & 0 & \multicolumn{1}{D{.}{\;\rightarrow\;}{-1}}{2^+_2.0^+_{gs}} &  75(3) &  70(6) & \\
       2342.1(1) & 1388.4(1) & 953.7 & \multicolumn{1}{D{.}{\;\rightarrow\;}{-1}}{0^+_1.2^+_1} &  188(4)  & 191(8)  & 95(7) \\
       2803.2(1) & 1849.4(1) & 953.7 & \multicolumn{1}{D{.}{\;\rightarrow\;}{-1}}{2^+_3.2^+_1} &  70(2)  & 72(6) & 13(5)\\
       2885.1(5) & 1931.3(5) & 953.7 & \multicolumn{1}{D{.}{\;\rightarrow\;}{-1}}{2^+_4.2^+_1} &  6.5(16)  &  & -0.5(41)\\
       2961.4(5) & 619.1(1) & 2342.1 & \multicolumn{1}{D{.}{\;\rightarrow\;}{-1}}{(1^+).0^+_1} &    10.8(14) &  & 14(4) \\
        & 1156.8(2) & 1804.5 & \multicolumn{1}{D{.}{\;\rightarrow\;}{-1}}{(1^+).2^+_2} &   17(2)  & 19(5) & \\
        & 2962.3(8) & 0 & \multicolumn{1}{D{.}{\;\rightarrow\;}{-1}}{(1^+).0^+_{gs}} &  5.1(16)  &  & \\
       3046.3(1) & 2092.5(1) & 953.7 & \multicolumn{1}{D{.}{\;\rightarrow\;}{-1}}{0^+_2.2^+_1} &  46(2)  & 12(5)\footnotemark[1]  & 45(2) \\
       3147.0(5) & 2193.2(8) & 953.7 & \multicolumn{1}{D{.}{\;\rightarrow\;}{-1}}{(2^+_5).2^+_1} &   2.2(9)  &  & 0.6(15) \\
       3180.9(3) & 295.5(14) & 2885.0 & \multicolumn{1}{D{.}{\;\rightarrow\;}{-1}}{1^{(+)}.2^+_4}   &  4.1(36) &  & 339(9) \\
        & 838.7(4) & 2342.1 & \multicolumn{1}{D{.}{\;\rightarrow\;}{-1}}{1^{(+)}.0^+_2}   &  7(3) &  & \\
        & 1376.2(3) & 1804.5 & \multicolumn{1}{D{.}{\;\rightarrow\;}{-1}}{1^{(+)}.2^+_2}   &  8.0(19) &  & \\
        & 2227.0(1) & 953.7 & \multicolumn{1}{D{.}{\;\rightarrow\;}{-1}}{1^{(+)}.2^+_1}   &  282(6) & 262(11) & \\
        & 3180.8(1) & 0 & \multicolumn{1}{D{.}{\;\rightarrow\;}{-1}}{1^{(+)}.0^+_{gs}}   &  37.4(17) & 42(5) & \\
       3339.8(5) & 2386.0(5) & 953.7 & \multicolumn{1}{D{.}{\;\rightarrow\;}{-1}}{(1^+).2^+_1} & 6.4(15)  &  & 4.0(19) \\
       3374.2(7) & 1031.4(2) & 2342.1 & \multicolumn{1}{D{.}{\;\rightarrow\;}{-1}}{(1^{\pm}).0^+_1} &  5.0(9)  & 6(5) & 61(3) \\
        & 1569.7(1) & 1804.5 & \multicolumn{1}{D{.}{\;\rightarrow\;}{-1}}{(1^{\pm}).2^+_2} &  37(2) & 30(5) &  \\
        & 3374.1(2) & 0 & \multicolumn{1}{D{.}{\;\rightarrow\;}{-1}}{(1^{\pm}).0^+_{gs}} &  21.5(15) & 18(4) &  \\
       3424.6(7) & 1619.3(1) & 1804.5 & \multicolumn{1}{D{.}{\;\rightarrow\;}{-1}}{(1^+).2^+_2} &  40(2) & 35(5) & 64(5) \\
        & 2471.4(5) & 953.7 & \multicolumn{1}{D{.}{\;\rightarrow\;}{-1}}{(1^+).2^+_1} &  12(5) &  & \\
        & 3425.3(3) & 0 & \multicolumn{1}{D{.}{\;\rightarrow\;}{-1}}{(1^+).0^+_{gs}} & 12(2) &  & \\
       3560.9(5) & 2607.1(5) & 953.7 & \multicolumn{1}{D{.}{\;\rightarrow\;}{-1}}{(1^+).2^+_1} &  3.9(14)  &  & 3.9(14) \\
       3690.5(9) & 2736.7(9) & 953.7 & \multicolumn{1}{D{.}{\;\rightarrow\;}{-1}}{(1^+).2^+_1} &  3.0(10) &  & 3.0(10) \\
       4021.7(5) & 1218.5(6) & 2803.1 & \multicolumn{1}{D{.}{\;\rightarrow\;}{-1}}{(1^+).2^+_3} &  13(3)  &  & 205(8) \\
        & 1680.4(4) & 2342.1 & \multicolumn{1}{D{.}{\;\rightarrow\;}{-1}}{(1^+).0^+_1} &  8(2) & 8(5) & \\
        & 3067.8(2) & 953.7 & \multicolumn{1}{D{.}{\;\rightarrow\;}{-1}}{(1^+).2^+_1} &   15.3(13) & 16(4) & \\
        & 4021.6(1) & 0 & \multicolumn{1}{D{.}{\;\rightarrow\;}{-1}}{(1^+).0^+_{gs}} &  169(6) & 149(10) & \\
       4322.4(9) & 1982.1(9) & 2342.1 & \multicolumn{1}{D{.}{\;\rightarrow\;}{-1}}{(1^+).0^+_1}  & 0.8(5)  &  & 16(3) \\
        & 2518.0(6) & 1804.5 & \multicolumn{1}{D{.}{\;\rightarrow\;}{-1}}{(1^+).2^+_2}  &  9.7(14) &  & \\
        & 3368.1(5) & 953.7 & \multicolumn{1}{D{.}{\;\rightarrow\;}{-1}}{(1^+).2^+_1}  &  5(2) &  & \\
       4447.9(4) & 1107.4(6) & 3339.7 & \multicolumn{1}{D{.}{\;\rightarrow\;}{-1}}{(1^+).(1^+)} & 0.9(4) & & 277(9) \\
        & 1644.6(2) & 2803.1 & \multicolumn{1}{D{.}{\;\rightarrow\;}{-1}}{(1^+).2^+_3} &  16(2) & 21(5) &  \\
        & 2105.4(1) & 2342.1 & \multicolumn{1}{D{.}{\;\rightarrow\;}{-1}}{(1^+).0^+_1} &  43(2) & 61(6) & \\
        & 2643.2(1) & 1804.5 & \multicolumn{1}{D{.}{\;\rightarrow\;}{-1}}{(1^+).2^+_2} &  36.4(16) & 26(5) & \\
        & 3493.9(1) & 953.7 & \multicolumn{1}{D{.}{\;\rightarrow\;}{-1}}{(1^+).2^+_1} &  64(2) & 58(6) & \\
        & 4447.7(1) & 0 & \multicolumn{1}{D{.}{\;\rightarrow\;}{-1}}{(1^+).0^+_{gs}} &  117(8) & 109(8) & \\
       4531.5(5) & 3577.6(5) & 953.7 & \multicolumn{1}{D{.}{\;\rightarrow\;}{-1}}{(1^+).2^+_1} & 3.1(12) & & 3.1(12) \\
       4894.4(5) & 1932.4(2) & 2961.3 & \multicolumn{1}{D{.}{\;\rightarrow\;}{-1}}{(1^+).(1^+)} & 19(2) & & 94(9) \\
        & 2091.5(4) & 2803.1 & \multicolumn{1}{D{.}{\;\rightarrow\;}{-1}}{(1^+).2^+_3} &  7(2) & 47(6)\footnotemark[1] & \\
        & 3089.6(6) & 1804.5 & \multicolumn{1}{D{.}{\;\rightarrow\;}{-1}}{(1^+).2^+_2} &  1.3(9) & 9(4) & \\
        & 3940.5(4) & 953.7 & \multicolumn{1}{D{.}{\;\rightarrow\;}{-1}}{(1^+).2^+_1} &  3.2(14) &  & \\
        & 4894.2(1) & 0 & \multicolumn{1}{D{.}{\;\rightarrow\;}{-1}}{(1^+).0^+_{gs}} & 64(8) & 42(5) & \\
       5210.8(4) & 1836.9(4) & 3374.1 & \multicolumn{1}{D{.}{\;\rightarrow\;}{-1}}{(1^+).(1^\pm)} & 2.4(10) &  & 139(11) \\
        & 2062.8(2) & 3146.9 & \multicolumn{1}{D{.}{\;\rightarrow\;}{-1}}{(1^+).(2^+_5)} & 1.6(12) &  & \\ 
        & 2164.4(5) & 3046.2 & \multicolumn{1}{D{.}{\;\rightarrow\;}{-1}}{(1^+).0^+_2} & 1.1(6) &  & \\
        & 2407.8(3) & 2803.1 & \multicolumn{1}{D{.}{\;\rightarrow\;}{-1}}{(1^+).2^+_3} & 11.7(15) & 13(4) & \\
        & 3405.8(5) & 1804.5 & \multicolumn{1}{D{.}{\;\rightarrow\;}{-1}}{(1^+).2^+_2} & 8.0(13)  &  & \\
        & 4256.9(1) & 953.7 & \multicolumn{1}{D{.}{\;\rightarrow\;}{-1}}{(1^+).2^+_1} & 40(2)  & 29(4) & \\
        & 5210.6(1) & 0 & \multicolumn{1}{D{.}{\;\rightarrow\;}{-1}}{(1^+).0^+_{gs}} & 62(10) & 51(6) & \\
       5394.3(10) & 2589.9(7) & 2803.1 & \multicolumn{1}{D{.}{\;\rightarrow\;}{-1}}{(1^+).2^+_3} & 0.8(4)   &  & 11(2) \\
        & 3051.6(7) & 2342.1 & \multicolumn{1}{D{.}{\;\rightarrow\;}{-1}}{(1^+).0^+_1} & 2.5(9)  &  & \\
        & 3589.6(4) & 1804.5 & \multicolumn{1}{D{.}{\;\rightarrow\;}{-1}}{(1^+).2^+_2} & 8(2) &  & \\
       5508.0(11) & 2166.7(4) & 3339.7 & \multicolumn{1}{D{.}{\;\rightarrow\;}{-1}}{(1^+).(1^+)} & 1.5(10)  &  & 18(2) \\
        & 2624.2(4) & 2885.0 & \multicolumn{1}{D{.}{\;\rightarrow\;}{-1}}{(1^+).2^+_4} & 2.9(11) &  & \\
        & 2704.4(3) & 2803.1 & \multicolumn{1}{D{.}{\;\rightarrow\;}{-1}}{(1^+).2^+_2} & 8.6(14) &  & \\
        & 3164.1(7) & 2342.5 & \multicolumn{1}{D{.}{\;\rightarrow\;}{-1}}{(1^+).0^+_1} & 1.7(8) &  & \\
        & 5508.0(7) & 0 & \multicolumn{1}{D{.}{\;\rightarrow\;}{-1}}{(1^+).0^+_{gs}} & 3.7(13) &  & \\
       5583.8(4) & 5583.5(4) & 0 & \multicolumn{1}{D{.}{\;\rightarrow\;}{-1}}{(1^+).0^+_{gs}} & 1.3(5) &  & 1.3(5) \\
       5919.6(6) & 3577.7(6) & 2342.1 & \multicolumn{1}{D{.}{\;\rightarrow\;}{-1}}{(1^+).0^+_1} & 1.1(7)  &  & 16(5)\\
        & 5919.2(5) & 0 & \multicolumn{1}{D{.}{\;\rightarrow\;}{-1}}{(1^+).0^+_{gs}} & 15(5) & 8(4) & 
\\
\end{longtable*}
\end{center}

\subsection{Gamma-Gamma Angular Correlations}

One of the main objectives of the current experiment was to definitively establish the location of the first-excited 0$^{+}$ state in $^{62}$Zn, which has a major effect on the calculation of the isospin symmetry breaking correction $\delta_{C1}$ in $^{62}$Ga superallowed decay~\cite{Towner2008}. The 2342-keV state was assumed to be the first-excited 0$^{+}$ state in $^{62}$Zn in the previous $^{62}$Ga superallowed branching ratio measurements reported in Refs.~\cite{Hyland2005,Finlay2008,Bey2008} and this assignment was supported by the $\gamma-\gamma$ angular correlation analysis reported in Ref.~\cite{Albers2010} over 10 angular correlation groups, as well as angular distribution measurements from $^{58}$Ni($^6$Li,$d$)$^{62}$Zn~\cite{Fulbright1977}, $^{64}$Zn(p,t)$^{62}$Zn~\cite{Hinrichs1974, Kusakari1972}. Subsequently, however, the 2342-keV state was assigned a spin-parity $J^\pi=2^{+}$ in Refs.~\cite{Leach2013,Leach2019} based on the measured triton angular distributions in the $^{64}$Zn($p,t$)$^{62}$Zn transfer reaction. In Refs.~\cite{Leach2013,Leach2019}, the 3046-keV state was assigned as the first-excited 0$^{+}$ state in $^{62}$Zn, which would have important implications for the configuration mixing component of the isospin symmetry breaking correction, $\delta_{C1}$, in $^{62}$Ga superallowed decay, changing the central value by almost a factor of two from $\delta_{C1} = 0.275(55)$~\cite{Towner2008} to $\delta_{C1} = 0.160(70)$~\cite{Leach2013}.

\begin{figure}[ht!]
\begin{center}
\includegraphics[width=\columnwidth, keepaspectratio]{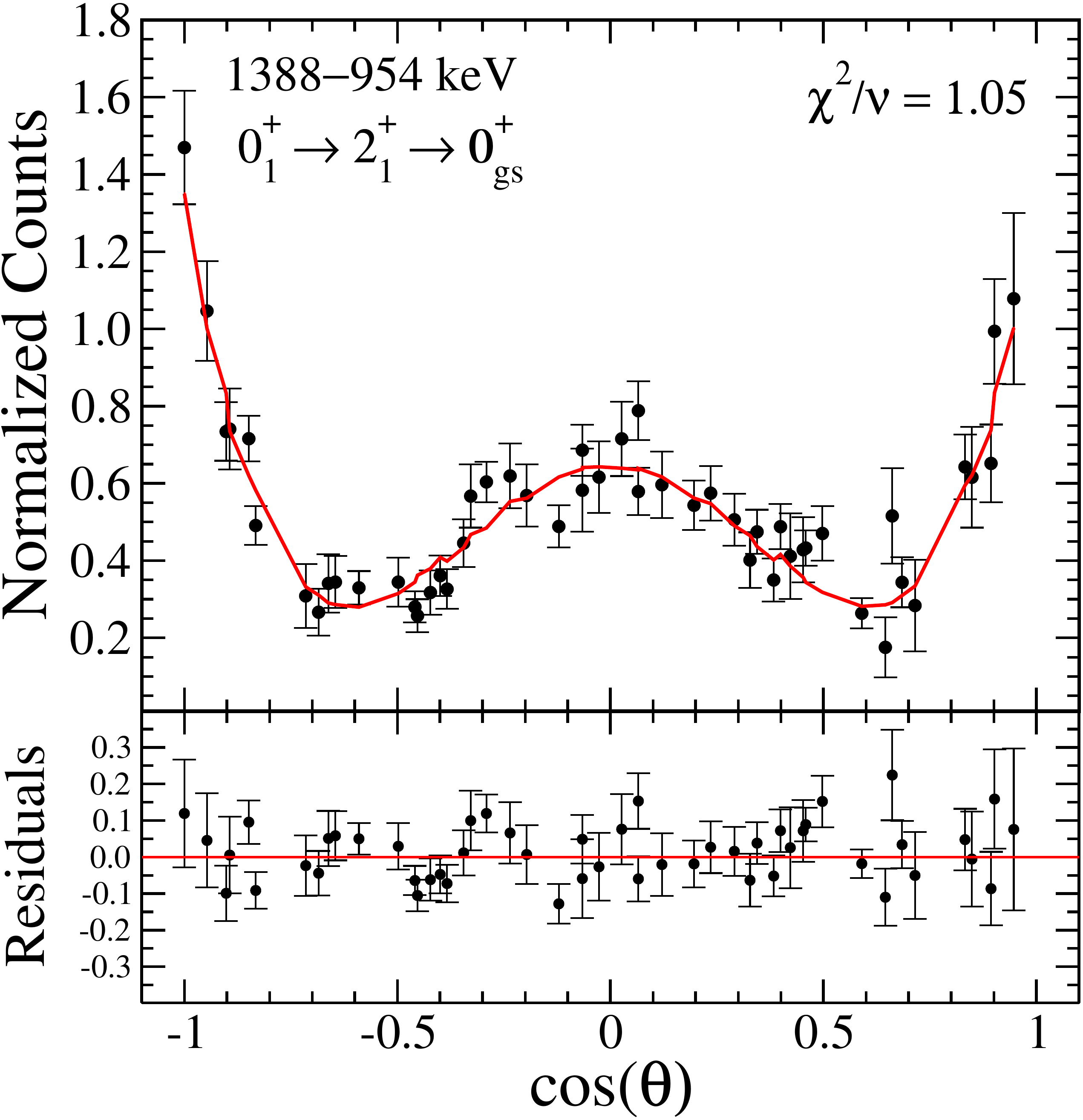}
\caption{(Colour online) Experimental data for the $\gamma-\gamma$ angular correlation between the 1388 and 954~keV $\gamma$-rays. The red line is not a fit to the data, but the result of a Geant4~\cite{Geant4} simulation of the expected $\gamma-\gamma$ angular correlation for a $0^{+}_{1}\rightarrow 2^{+}_{1}\rightarrow 0^{+}_{gs}$ cascade. }
\label{fig:020AC}
\end{center}
\end{figure}

\begin{figure}[ht!]
\begin{center}
\includegraphics[width=\columnwidth, keepaspectratio]{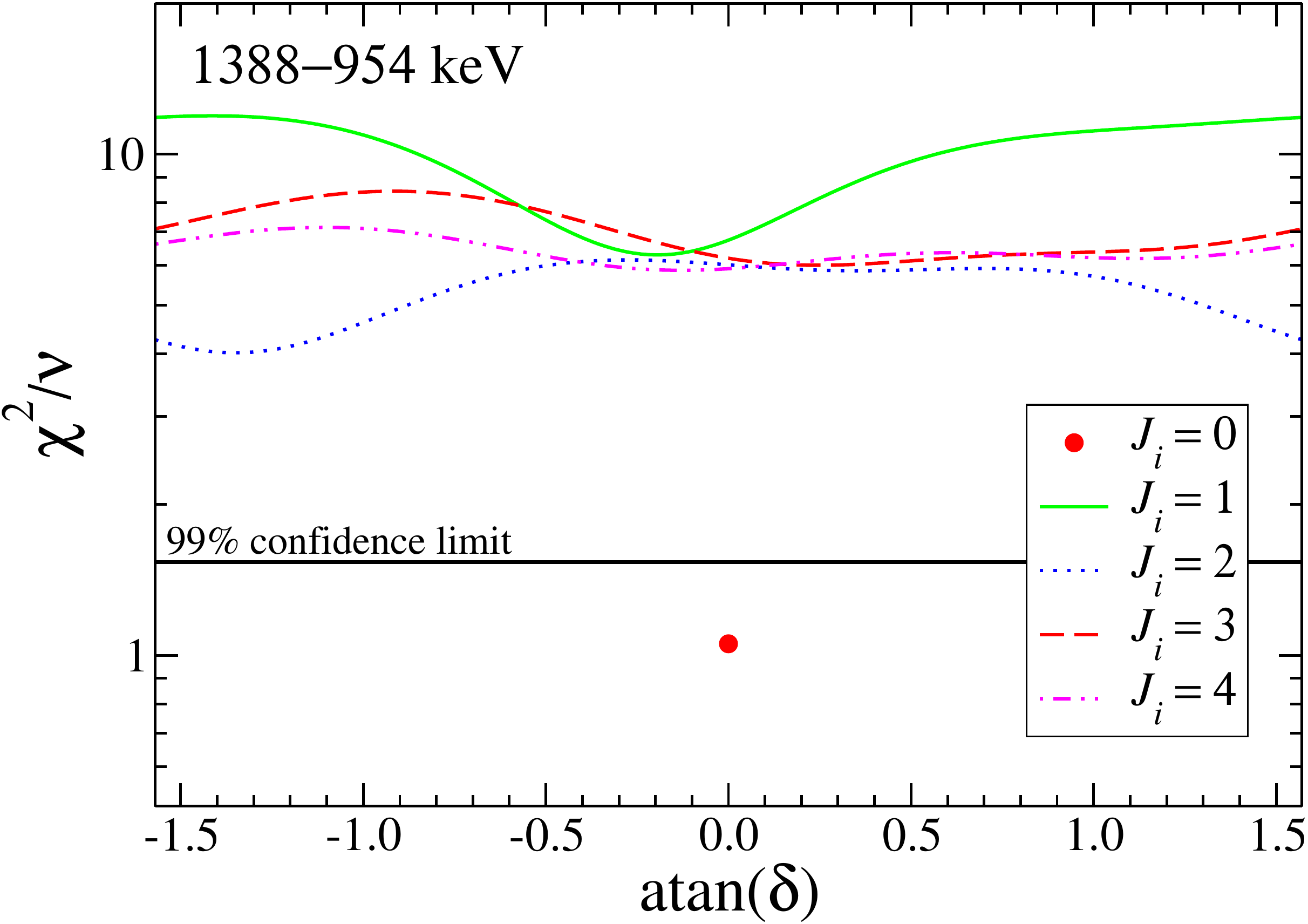}
\caption{(Colour online) $\chi^{2}/\nu$ versus $\arctan$ of the mixing ratio, $\delta$, for the 1388-keV $\gamma$~ray in the 1388--954-keV $\gamma-\gamma$ cascade under different spin assumptions for the 2342-keV state in $^{62}$Zn. For this cascade the $\chi^{2}/\nu$ minimum allows a definitive spin assignment of $J = 0$ for the 2342-keV state.}
\label{fig:020ACChi}
\end{center}
\end{figure}

To definitively establish the spin of the 2342-keV state in $^{62}$Zn, a $\gamma-\gamma$ angular correlation analysis was performed for the 1388--954-keV $\gamma$-ray cascade following $^{62}$Ga $\beta$ decay using the 2048 pairs of HPGe crystals in the GRIFFIN spectrometer to define 52 distinct opening angles between detectors as described in Ref.~\cite{Smith2019}. The ``Method 1'' approach described in Ref.~\cite{Smith2019} was used where high-statistics Geant4~\cite{Geant4} simulations were performed that included the precise geometry of the GRIFFIN spectrometer and direct comparisons were made between simulated and experimental $\gamma-\gamma$ angular correlation data. For the 1388--954-keV cascade this comparison is shown in Figure~\ref{fig:020AC} and is in excellent agreement with the expectation for a $0^{+}\rightarrow 2^{+} \rightarrow 0^{+}$ cascade. In order to explore the viability of other possible spin assignments for the 2342-keV state, the $\chi^{2}/\nu$ from the comparison of the Geant4 simulations to the experimental $\gamma-\gamma$ angular correlation data were plotted as a function of the assumed mixing ratio, $\delta$, for the 1388-keV transition, as shown in Fig.~\ref{fig:020ACChi}. It is clear from this analysis that the only possible spin assignment is $J = 0$, with a $\chi^2/\nu$ of 1.05, while all other assumed spin values for the 2342-keV state give $\chi^{2}/\nu > 4$. The minimum $\chi^{2}/\nu$ under the assumption of a 2$^{+}$ assignment is found to be 4.1, at a mixing ratio of $\delta = -4.3$. From this analysis, the $J^\pi=2^+$ assignment proposed in Refs.~\cite{Leach2013,Leach2019} can be excluded with 99.9$\%$ confidence. The population of the 2342-keV state in the $^{62}$Zn($p,t$)$^{62}$Zn transfer reaction, as well as its direct $\beta$ feeding in $^{62}$Ga $\beta$ decay (see below) indicates that the state has natural parity. We therefore assign the spin and parity of the 2342~keV excited state to be $0^{+}$, in agreement with the previous transfer reaction measurements of Refs.~\cite{ Kusakari1972, Hinrichs1974, Fulbright1977} and the previous $\gamma-\gamma$ angular correlation measurement of Ref.~\cite{Albers2010}, superseding the 2$^{+}$ assignment proposed in our transfer reaction work described in Refs.~\cite{Leach2013,Leach2019}.  

\begin{figure}[ht!]
\begin{center}
\includegraphics[width=\columnwidth, keepaspectratio]{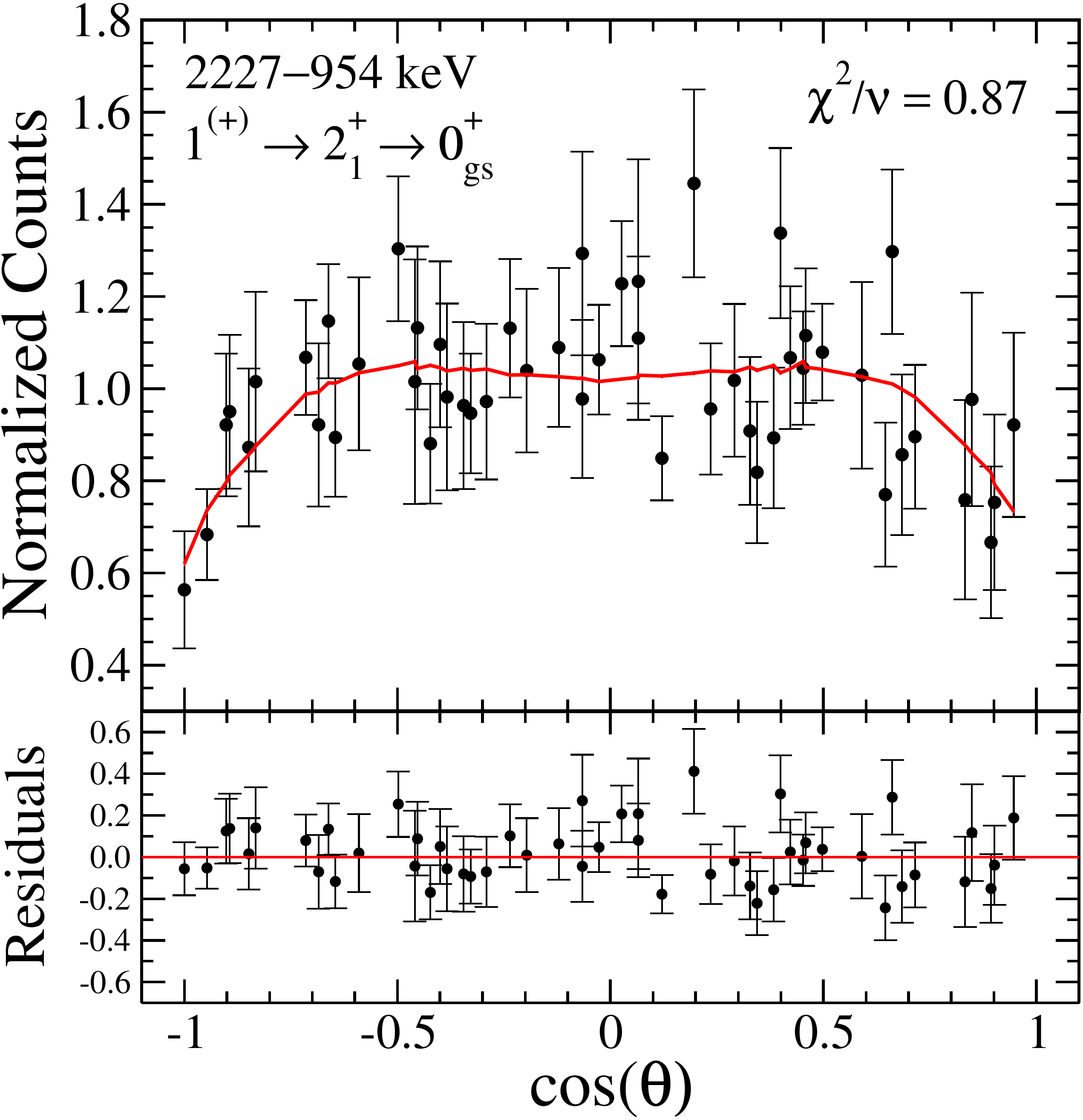}
\caption{(Colour online) The measured $\gamma-\gamma$ angular correlation between the 2227-keV and 954-keV $\gamma$~rays is shown by the data points with error bars. The red line is not a fit to the data, but the result of a Geant4~\cite{Geant4} simulation of the expected $\gamma-\gamma$ angular correlation for a $1^{(+)}\rightarrow 2^{+}_{1}\rightarrow 0^{+}_{gs}$ cascade.}
\label{fig:120AC}
\end{center}
\end{figure}

\begin{figure}[ht!]
\begin{center}
\includegraphics[width=\columnwidth, keepaspectratio]{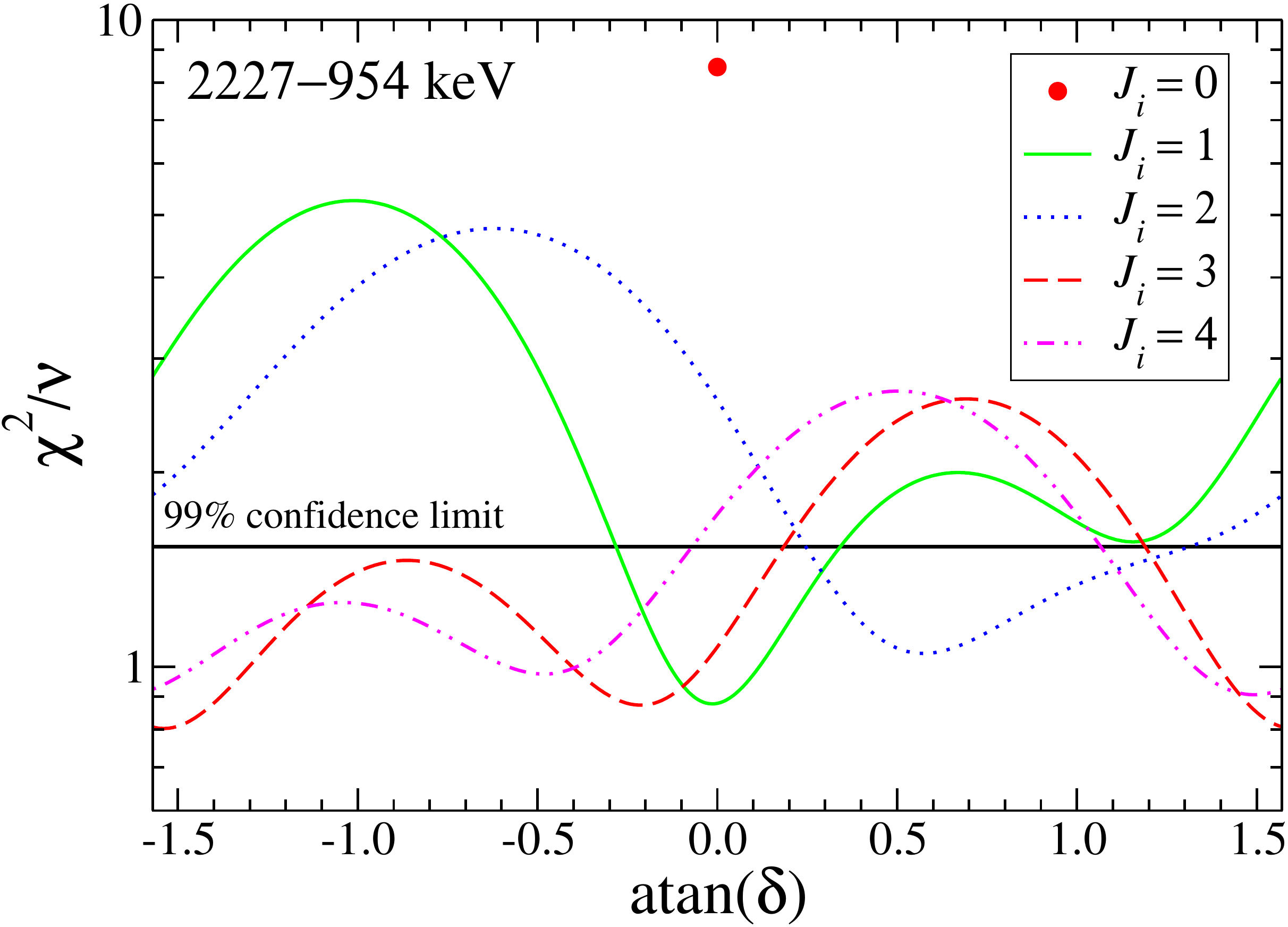}
\caption{(Colour online) $\chi^{2}/\nu$ versus $\arctan$ of the mixing ratio, $\delta$, for the 2227-keV $\gamma$~ray in the 2227--954-keV $\gamma-\gamma$ cascade under different spin assumptions for the 3181-keV state in $^{62}$Zn. For this  cascade the $\chi^{2}/\nu$ does not allow a definitive spin assignment for the 3181-keV state, but combined with $\beta$ feeding information an assignment of $J^{\pi} = 1^{(+)}$ (green, solid curve) with a mixing ratio of $\delta = -0.01(8)$ is adopted (see text for details).}
\label{fig:120ACChi}
\end{center}
\end{figure}

An analysis of the $\gamma-\gamma$ angular correlation for the 2227--954-keV cascade was also performed. The 3181-keV excited state of $^{62}$Zn was previously assigned tentatively as $J^\pi=(1^{+})$~\cite{Hyland2005,Finlay2008}, thus suggesting a $(1^{+})\rightarrow2^{+}\rightarrow0^{+}$ cascade which, as shown in Fig.~\ref{fig:120AC}, is entirely consistent with the $\gamma-\gamma$ angular correlation measured in the current work (with $\chi^2/\nu = 0.87$). As shown in Fig.~\ref{fig:120ACChi}, the spin of the 3181-keV state could not be uniquely assigned based only on the measured $\gamma-\gamma$ angular correlation. However, a 0$^{+}$ assignment for the 3181-keV state could be excluded, consistent with the observed direct $\gamma$ decay to the 0$^{+}$ ground state. If we further consider the $\beta^{+}$ decay selection rules, we note that the 3181-keV state has the largest $I_{out}-I_{in}$ value of any level observed in the current work at 339(9)~ppm, indicating that this state is fed directly by the $\beta^{+}$ decay of $^{62}$Ga. This suggests a $J^\pi=1^+$ level populated by allowed Gamow-Teller $\beta$-decay of the $J^\pi=0^+$ ground state of $^{62}$Ga. 
The $J=3$ and 4 spin assignments allowed by the $\gamma-\gamma$ angular correlation analysis can be excluded as their direct $\beta$ feeding from the $J^\pi=0^+$ parent would require, at minimum, second- and third-forbidden $\beta$ decays and would be completely negligible. A $J^\pi=2^+$ assignment, as proposed in Ref.~\cite{Leach2019}, requiring a second-forbidden $\beta$ decay, can similarly be excluded while a $J^\pi=2^-$ assignment would require a first-forbidden unique $\beta$ decay followed by an improbable $M2$ $\gamma$ decay to the ground state competing with an $E1$ decay to the $2^{+}_{1}$ state. The $\log(ft)$ value for this state was determined to be 5.95(2), which favours the assignment of this state as 1$^{+}$ although it does not completely exclude the possibility of a 1$^{-}$ state fed by a strong first forbidden $\beta$ decay for  which
the recommended cutoff is $\log(ft) > 5.9$~\cite{Singh1998}. Combining this information, $J^{\pi} = $ 1$^{(+)}$ is assigned to the 3181-keV state. The minimum $\chi^{2}/\nu$ from the $\gamma-\gamma$ angular correlation analysis with this spin assignment (green, solid curve of Fig.~\ref{fig:120ACChi}), yields a mixing ratio of $\delta = -0.01(8)$ for the 2227-keV $1^{(+)}\rightarrow 2^+_1$ transition, indicating that it is $M1$ dominated with little to no $E2$ strength.

\begin{figure}[ht!]
\begin{center}
\includegraphics[width=\columnwidth, keepaspectratio]{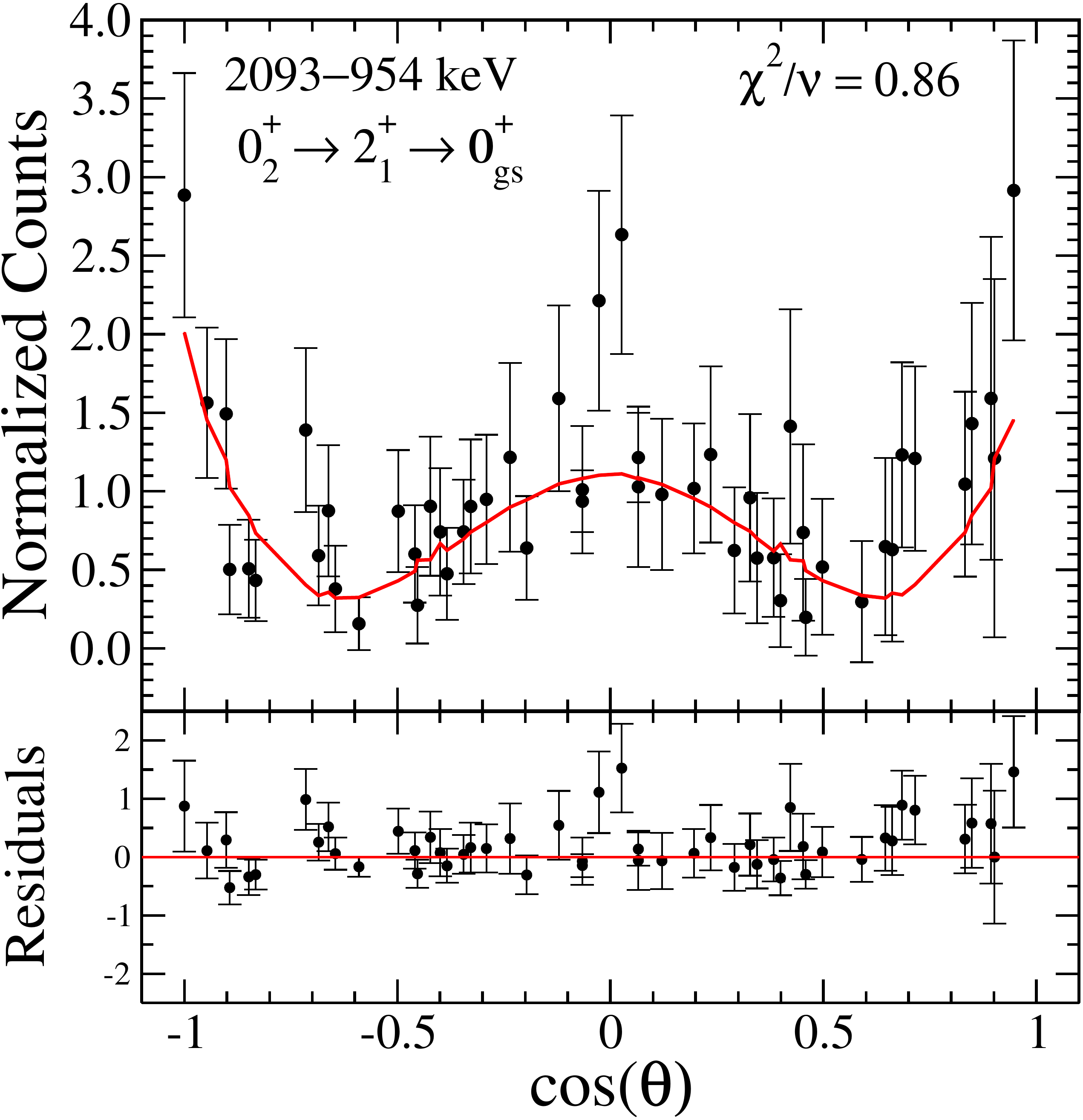}
\caption{(Colour online) The measured $\gamma-\gamma$ angular correlation between the 2093-keV and 954-keV $\gamma$-rays is shown by the data points with error bars. The red line is not a fit to the data, but the result of a Geant4~\cite{Geant4} simulation of the expected $\gamma-\gamma$ angular correlation for the $0^{+}_{2}\rightarrow 2^{+}_{1}\rightarrow 0^{+}_{gs}$ cascade. }
\label{fig:220AC}
\end{center}
\end{figure}

\begin{figure}[ht!]
\begin{center}
\includegraphics[width=\columnwidth, keepaspectratio]{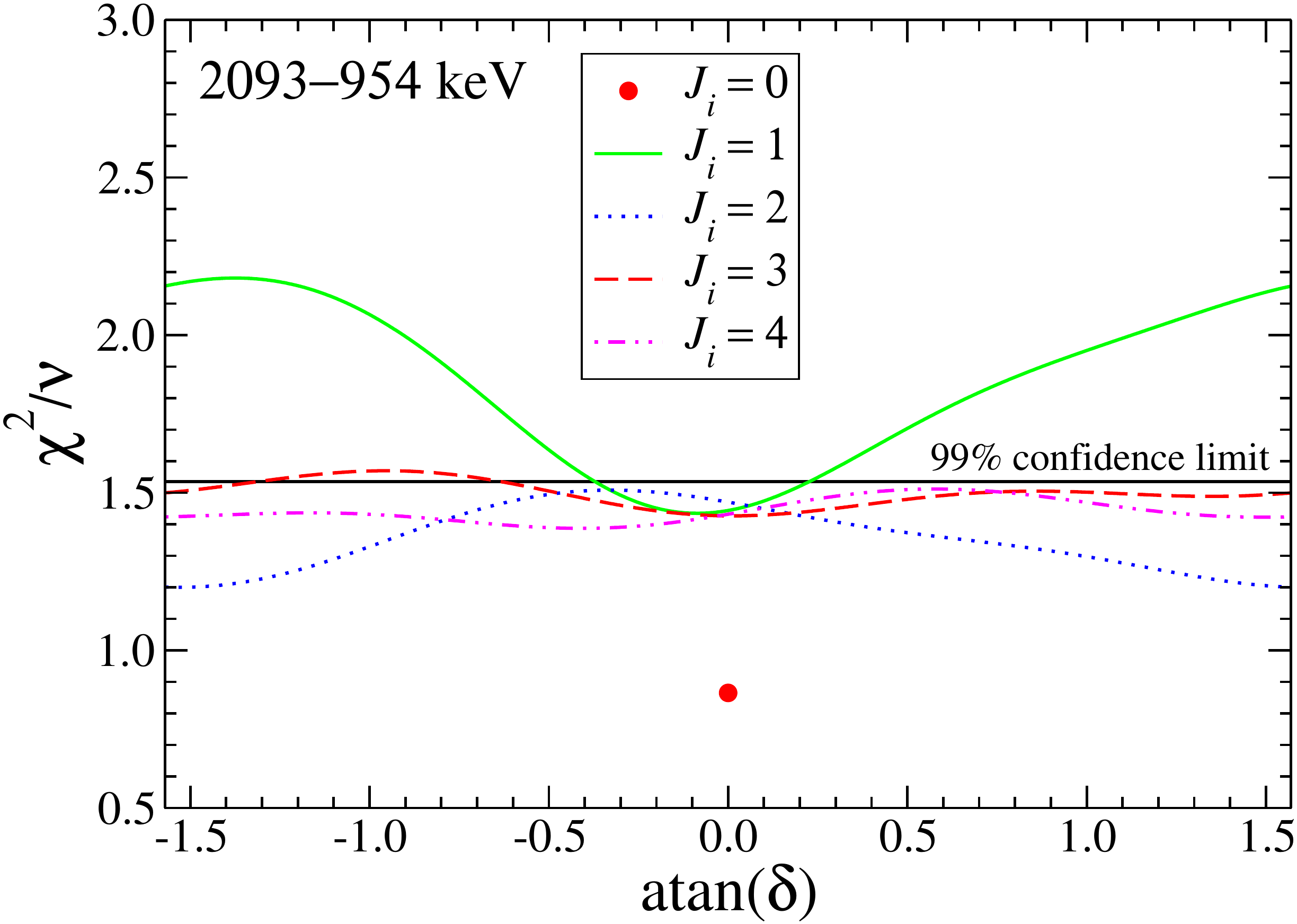}
\caption{(Colour online) $\chi^{2}/\nu$ versus $\arctan$ of the mixing ratio, $\delta$, for the 2093-keV $\gamma$~ray in the 2093--954-keV $\gamma-\gamma$ cascade under different spin assumptions for the 3046-keV state in $^{62}$Zn. For this cascade the $\chi^{2}/\nu$ does not allow a definitive spin assignment for the 3046-keV state, but is consistent with the tentative $J^{\pi} = 0^{+}$ assignment. }
\label{fig:220ACChi}
\end{center}
\end{figure}

An analysis of the $\gamma-\gamma$ angular correlation for the 2093--954-keV cascade originating from the 3046-keV excited state was also performed. This state was tentatively assigned as the second-excited $J^\pi=0^+$ state in $^{62}$Zn in Ref.~\cite{Finlay2008} and also identified as a $J^\pi=0^+$ state from the angular distribution of the $^{64}$Zn($p,t$)$^{62}$Zn transfer reaction in Refs.~\cite{Leach2013,Leach2019}. For the $\gamma-\gamma$ angular correlation measurement presented in the current work, the 2092/2093-keV doublet required additional consideration. The 2092-keV transition, with an intensity of 7(2)~ppm, originates from the excited level at 4894~keV and is assigned as a (1$^{+}$)$\rightarrow 2^{+}_{3}$ transition. The 2093-keV transition of interest from the excited level at 3046~keV to the 954-keV $2^+_1$ level has an intensity of 46(2)~ppm and thus dominated the overall intensity of the doublet. The contamination from the weaker 2092-keV transition, which is also in coincidence with the 954~keV, must nonetheless be considered. Assuming that the 2092-keV transition is also dominantly $M1$, similar to what was observed for the 2227-keV transition from the 3181-keV $1^{(+)}$ level, a normalized fraction of the $\gamma-\gamma$ angular correlation measured for the 2227--954-keV cascade was subtracted from the $\gamma-\gamma$ angular correlation observed for the 2092/2093-keV doublet--954-keV cascade. The purpose of this was to remove the contribution from the 2092-keV transition and extract only the 2093--954-keV correlation. The results of this analysis is shown in Fig.~\ref{fig:220AC} and is fully consistent ($\chi^2/\nu=0.86$) with the expected $\gamma-\gamma$ angular correlation for a $0^+\rightarrow 2^+ \rightarrow 0^+$ cascade. As shown in Fig.~\ref{fig:220ACChi}, the spin of the 3046-keV state could not be definitively assigned based only on the $\gamma-\gamma$ angular correlations measured in the current work, but combined with the transfer reaction angular distribution reported in Refs.~\cite{Leach2013,Leach2019} provide confidence that the 3046-keV level is indeed the second excited $J^\pi=0^+$ state in $^{62}$Zn.

Finally, an analysis to confirm the mixing ratio of the 851-keV transition from the 1805-keV 2$^+_2$ level was also performed using the 851--954-keV $\gamma-\gamma$ cascade. The minimum $\chi^2/\nu = 0.82$ was obtained for $\delta = -5.1^{+2.9}_{-3.4}$ for the 851-keV transition, in exact agreement with Ref.~\cite{Ward1981}, and also excellent agreement with, although less precise than, the value of $\delta=-3.62^{+0.63}_{-1.03}$ previously determined in the work of Ref.~\cite{Albers2010}. 

\section{Discussion}
\label{sec:discussion}
\subsection{Superallowed Branching Ratio}

The isobaric analogue state of the $\beta$-decaying $J^{\pi} = 0^{+}$ ground state of $^{62}$Ga is the ground state of the daughter $^{62}$Zn. Therefore, the superallowed $\beta$ branching ratio must be measured indirectly by determining the sum of the intensities of all $\gamma$-ray transitions feeding the $^{62}$Zn ground state following $^{62}$Ga $\beta$ decay and subtracting this sum from unity. The intensities of all $\gamma$ rays identified in the current work are listed in Table~\ref{tab:gione} and the sum of the intensities of the observed direct ground state $\gamma$-ray transitions was $I^{obs}_{gs} = 1420(23)$~ppm. 

As discussed in Ref.~\cite{Hardy2002}, $^{62}$Zn is, however, predicted to have over one hundred $1^{+}$ states accessible via nonsuperallowed $\beta$ branches of $^{62}$Ga, many of which will go unobserved due to their weak population and the subsequent fragmentation of their $\gamma$ decays toward the ground state, a particular application of the so-called ``Pandemonium'' effect~\cite{Hardy1977} to the case of high $Q$-value superallowed $\beta$ decays. Although each of these weakly populated 1$^{+}$ states individually contributes negligibly to the nonsuperallowed branching ratio, summing over all such states could, in principle, make a non-negligable contribution given the large $Q$-value for $^{62}$Ga $\beta$ decay ($Q_{EC} = 9181.4(1)$~keV~\cite{Erornen2006}). In order to correct for the unobserved $\gamma$ decay to the ground state from these weakly populated 1$^{+}$ states, we follow the procedures for high-precision branching ratio measurements developed by our collaboration in Refs.~\cite{Hyland2005,Finlay2008,Dunlop2013} and combine the measured $\gamma$-ray intensities feeding and decaying from each observed level reported in Table~\ref{tab:gione} with calculations of relative $\gamma$-ray branching ratios averaged over the large number of predicted 1$^{+}$ states.

The final column of Table~\ref{tab:gione} gives the difference in the total intensity of the $\gamma$ rays emitted from a level and the total intensity of the $\gamma$ rays observed feeding that level. For 0$^{+}$ and 1$^{+}$ states, this difference represents a combination of both direct $\beta$ feeding (nonanalogue Fermi for the excited 0$^+$ states and allowed Gamow-Teller for the 1$^+$ states) of these states and the possibility of unobserved $\gamma$-ray feeding from higher-lying states. For the 2$^+$ states, however, this difference must be entirely due to unobserved $\gamma$-ray feeding from above as direct population of these states would require a second-forbidden $\beta$ decay and is expected to be entirely negligible. In the current work, the total unobserved $\gamma$-ray feeding into the $2^+_1$, $2^+_2$, $2^+_3$, $2^+_4$ and $2^+_5$ levels was thus measured to be $I'_{2^{+}} = 11(19)$~ppm. This result is to be compared, for example, with the value of $I_{2^{+}}' = 205(29)$~ppm reported in Ref.~\cite{Finlay2008} for the first three excited 2$^{+}$ states (as the 2$^{+}_{4}$ and $2^{+}_{5}$ states were not observed in that work). The high efficiency of the GRIFFIN spectrometer and, in particular, its high $\gamma-\gamma$ coincidence efficiency is thus seen to have reduced the unobserved $\gamma$-ray feeding to the low-lying 2$^{+}$ states by more than an order of magnitude to the point where it is now consistent with zero at the 1$\sigma$ level. As essentially all of the $\gamma$-ray feeding to the low-lying 2$^{+}$ levels of $^{62}$Zn is conclusively demonstrated to have been observed in the current work, there is no reason to expect that the same is not true for the low-lying 0$^{+}$ states, including the ground state, representing a major advance in controlling the Pandemonium problem~\cite{Hardy2002, Hardy1977} through high-efficiency, high-resolution $\gamma$-ray spectroscopy.

To quantify the above statement, we follow the procedures of Refs.~\cite{Hyland2005,Finlay2008,Dunlop2013}, and define
\begin{equation}
\label{eq:Bgs}
\overline{B}_{gs} = \frac{I'_{gs}}{I'_{gs}+I'_{2^+}}
\end{equation}
where $I'_{gs}$ is the unobserved $\gamma$-ray feeding to the ground state that we are seeking to quantify and $\overline{B}_{gs}$ represents the ratio of the $\gamma$-ray intensity, averaged over the large number of unobserved weakly populated $1^{+}$ levels, that proceeds to the ground state without passing through one of the low-lying 2$^{+}_{1}$, 2$^{+}_{2}$, 2$^{+}_{3}$, 2$^{+}_{4}$ or 2$^{+}_{5}$ collector states, to the total $\gamma$-decay intensity ($I'_{gs} + I'_{2^{+}}$) from these levels. To estimate the value of the average (weighted by $\beta$ feeding) ratio $\overline{B}_{gs}$, we use the shell-model calculations of $^{62}$Zn $\gamma$-decay following $^{62}$Ga $\beta$ decay described in detail in Ref.~\cite{Finlay2008}, which led to the adoption of $\overline{B}_{gs} = 0.20(20)$ with a very conservative $\pm 100\%$ uncertainty assigned to this theoretical value. This value of $\overline{B}_{gs}$, combined with the measured $I'_{2^{+}} = 11(19)$~ppm, yields the unobserved ground state $\gamma$ intensity as:

\begin{equation}
\label{eq:Igs}
I'_{gs} = I'_{2^{+}}\left(\frac{\overline{B}_{gs}}{1- \overline{B}_{gs}}\right) = 3^{+17}_{-3} \text{ ppm.}
\end{equation}

Combining this result with the observed ground state feeding $\gamma$-ray intensity of $I^{obs}_{gs} = 1420(23)$~ppm gives a nonsuperallowed branching ratio of 1423$^{+29}_{-23}$~ppm and a final superallowed branching ratio for $^{62}$Ga $\beta$ decay of 99.8577$^{+0.0023}_{-0.0029}$\%. A summary of the individual contributions to the uncertainty in the superallowed branching ratio is given in Table~\ref{tab:uncert}. Prior to the current work, the adopted world-average for the $^{62}$Ga superallowed branching ratio was $99.862\pm0.011\%$, based on a weighted average of 99.858(8)\% from Ref.~\cite{Finlay2008} and 99.893(24)\% from Ref.~\cite{Bey2008} with the uncertainty scaled by a factor of 1.4 to account for the inconsistency between the two results. The result from the current work is consistent with, but approximately a factor of 4 more precise than this previous world average.

\begin{table}[ht!]
\caption{Contributions to the uncertainty in determining the branching ratio for the superallowed $\beta$ decay of $^{62}$Ga. The final contribution, $I'_{gs}$, represents the unobserved ground state feeding and includes contributions from both the experimentally determined unobserved $\gamma$ feeding to the $2^+_{1-5}$ states, $I'_{2^+}$, and the theoretical calculation of the ground state branching ratio from the unobserved 1$^+$ states, $\overline{B}_{gs}$, as defined in Eqs.~\ref{eq:Bgs} and \ref{eq:Igs}.}
  \centering
  \label{tab:uncert}
  \begin{tabular}{{lc}}
    \hline
    \hline
      Source & Uncertainty\\
        \hline
  $\gamma$ Counting Statistics & $\pm$0.00075\\ 
  $\gamma$ Efficiency Calibration & $\pm$0.00213\\ 
  $\gamma$ Summing Corrections & $\pm$0.00018\\ 
  $\beta$ Counting Statistics/Contaminants/Dead Time & $\pm$0.00006\\ 
  \multirow{2}{*}{Unobserved Ground State $\gamma$ Feeding ($I'_{gs}$)} & $+0.00028$\\ 
  &$-0.00173$\\ 
  \hline
    \multirow{2}{*}{Total Superallowed BR Uncertainty} & $+0.00226$\\ 
    & $-0.00285$\\ \hline 
    \bottomrule
  \end{tabular}
\end{table}

Although the current experiment is statistically independent from, and employed a different experimental facility than, the previous work by our collaboration reported in Ref.~\cite{Finlay2008}, both analyses relied on the same theoretical calculations and analysis procedures to estimate $\overline{B}_{gs} = 0.20(20)$ in order to determine the unobserved ground state $\gamma$ feeding. We therefore do not take a weighted average with the results of Ref.~\cite{Finlay2008}, but supersede our previous work with the significantly more precise value reported in the current work. As the current result is an order of magnitude more precise than all other measurements of the $^{62}$Ga superallowed branching ratio reported to date, we follow the analysis procedures of Ref.~\cite{Hardy2015} and adopt the current high-precision result of 99.8577$^{+0.0023}_{-0.0029}$\% for the $^{62}$Ga superallowed Fermi $\beta$ decay branching ratio.

\subsection{Excited 0$^{+}$ States and Isospin Symmetry Breaking Corrections}
\label{sec:isospin}

As noted in the Introduction in relation to Eqs.~\ref{eq:eqThree} and \ref{eq:eqFour}, the individual components $\delta^{n}_{C1}$ of the isospin mixing correction $\delta_{C1}$ can, in principle, be measured experimentally for excited 0$^{+}$ states within the $\beta^{+}/EC$ decay $Q$-value window. Such state-by-state comparisons of theoretical and experimental $\delta_{C1}^{n}$ values provide particularly stringent tests of a theoretical model's ability to describe isospin symmetry breaking effects in superallowed Fermi $\beta$ decay. 

\begin{table*}[ht!]
\caption{Measured nonanalogue Fermi $\beta$ branching ratios ($B_{n}$) to excited 0$^{+}$ states in $^{62}$Zn and the corresponding isospin mixing components $\delta^{n}_{C1}$. The theoretical values of $\delta^{n}_{C1}$ in the final columns are not scaled to the square of the ratio of the theoretical and experimental excitation energies. For example, the excitation energy for the first excited 0$^{+}$ state in the MSDI calculation~\cite{Towner2002} is 2.263~MeV, for the GXPF1 calculation~\cite{Towner2008} is 2.321~MeV, and for the FPVH and FPD6* calculations~\cite{Ormand1995} is 2.33~MeV, compared to the experimental value of 2.3421(1)~MeV.}
  \centering
  \label{tab:Limits}
  \begin{tabular}{c*{9}{lcc|c|llll}}
    \hline
    \hline
      &  &  &  & \multicolumn{5}{c}{$\delta_{C1}^{n}(\%)$}\\ \cline{5-9}
     Level & \multicolumn{1}{c}{E$_{level}$} & \multicolumn{1}{c}{$B_{n}$}  & $f_{0}/f_{n}$ & Experiment & \multicolumn{5}{c}{Theory}\\
      & \multicolumn{1}{c}{(keV)} & (ppm) &  & \multicolumn{1}{c|}{(This work)} & MSDI & GXPF1 & FPVH & FPD6*\\
        \hline
  0$^{+}_{1}$ & 2342.1(1) &  \multicolumn{1}{S[table-format = 3.2]}{97(10)} & \multicolumn{1}{S[table-format = 2.2]|}{4.91} & \multicolumn{1}{S[table-format = 3.5]|}{0.048(5)} & 0.089 & 0.160 & 0.079 & 0.169 \\        
  0$^{+}_{2}$ & 3046.2(1) & \multicolumn{1}{S[table-format = 3.2]}{45(3)} & \multicolumn{1}{S[table-format = 2.2]|}{8.93} & \multicolumn{1}{S[table-format = 3.5]|}{0.040(3)} & 0.006 & 0.010 & & & \\
  0$^{+}_{3}$ & 3862(2)\footnote[1]{The excitation energies of states not observed in the current work are taken from Ref.~\cite{Leach2013}} & \multicolumn{1}{S[table-format = 3.2]}{{$\leq$}2.2} & \multicolumn{1}{S[table-format = 2.2]|}{19.9} & \multicolumn{1}{S[table-format = 3.5]|}{{$\leq$}0.004} & 0.199 & 0.001 & & \\
  0$^{+}_{4}$ & 3936(6)\footnotemark[1] & \multicolumn{1}{S[table-format = 3.2]}{{$\leq$}1.3} & \multicolumn{1}{S[table-format = 2.2]|}{21.6} & \multicolumn{1}{S[table-format = 3.5]|}{{$\leq$}0.003} & 0.000 & 0.022 & & \\
  0$^{+}_{5}$ & 4552(9)\footnotemark[1] & \multicolumn{1}{S[table-format = 3.2]}{{$\leq$}2.4} & \multicolumn{1}{S[table-format = 2.2]|}{44.4}  & \multicolumn{1}{S[table-format = 3.5]|}{{$\leq$}0.011} &  &  &  &\\
  \hline
   &  &  &  {Total $\delta_{C1}$}& \multicolumn{1}{S[table-format = 3.5]|}{0.110(9)\footnote[2]{Obtained by assuming $\delta_{C1}^1$ and $\delta_{C2}^2$ account for 80(3)\% of the total strength; see text for discussion.}}  &  0.350 & 0.221 & 0.286 & 0.379 \\
  \hline
    \bottomrule
  \end{tabular}
\end{table*}

For the case of $^{62}$Ga superallowed decay, the currently adopted isospin-mixing correction $\delta_{C1} = 0.275(55)\%$~\cite{Hardy2015}, results from the average, and range, of two shell-model calculations~\cite{Towner2008}, the first employing the modified surface delta interaction (MSDI) of Ref.~\cite{Koops1977} yielding $\delta_{C1} = 0.329\%$, and the second with the GXPF1 interaction of Refs.~\cite{Honma2002, Honma2004} yielding $\delta_{C1} = 0.219\%$. In each case, the quoted $\delta_{C1}$ values have been scaled by the square of the ratio of the theoretical and experimental excitation energies of the first excited 0$^{+}$ state as described in Ref.~\cite{Towner2008}. The MSDI calculation, shown in detail in Fig.~8 of Ref.~\cite{Finlay2008}, provides a good description of the low-lying excited state spectrum of $^{62}$Zn, but has the somewhat atypical feature that the largest isospin-mixing component is predicted with the third-excited 0$^{+}$ state calculated at 3.072~MeV excitation energy with  $\delta_{C1}^{3} = 0.199\%$. A smaller mixing component of $\delta_{C1}^{1} = 0.089\%$ is calculated for the first-excited 0$^{+}$ state at 2.263~MeV excitation energy and $\delta_{C1}^{2} = 0.006\%$ is predicted for the second-excited 0$^{+}$ calculated at 2.885~MeV excitation energy. In this calculation, the first three excited 0$^{+}$ states account for 84\% of the total $\delta_{C1} = 0.350\%$, with the remainder resulting from smaller $\delta_{C1}^{n}$ components for more highly excited 0$^{+}$ states with $n > 3$. In the GXPF1 calculation, the more typical result of mixing primarily with the first-excited 0$^{+}$ state is obtained, with $\delta_{C1}^{1} = 0.160\%$ accounting for 72$\%$ of the total $\delta_{C1}$.

These theoretical predictions can be confronted with the nonanalogue Fermi $\beta$ decay branches measured in the current experiment. With the $\gamma-\gamma$ angular correlation measurements reported in Section~\ref{sec:results} conclusively establishing the 2342-keV excited state as a 0$^{+}$ state, the measured $I_{out}-I_{in}$ for this state of 95(7)~ppm sets an upper limit on the direct $\beta$ feeding of the level through a nonanalogue Fermi $\beta$ decay branch of $^{62}$Ga. While 9 separate $\gamma$-ray transitions with a total intensity of 93(7)~ppm were observed feeding the 2342-keV level from higher-lying (1$^{+}$) states, the possibility remains that weak $\gamma$-ray transitions from other 1$^{+}$ states below the level of sensitivity achieved in the current experiment also contributed to the measured $I_{out}-I_{in}$ of 95(7)~ppm for the 2342-keV level. To set a limit on such transitions from the other (1$^{+}$) states observed in the current work, a gate was set on the 1388-keV $\gamma$-ray depopulating the 2342-keV level and peaks with fixed widths and centroids were fit at the locations that would correspond to the feeding transitions. Similarly, a gate set on the 2106-keV transition feeding the 2342-keV level was used to set a limit on a possible 537-keV transition from this 0$^{+}$ state to the 2$^{+}_{2}$ level at 1805~keV excitation energy. The combination of all such unobserved transitions from states observed in the current work made a contribution of 4(8)~ppm to the $I_{out}-I_{in}$ for the 2342~keV level. Finally, we must also consider feeding of the 2342-keV level from unobserved high-lying (1$^{+}$) states. To quantify the effect, we again turn to the shell-model calculations, which indicate that the predicted $\gamma$ feeding into the 2342-keV 0$_{1}^{+}$ state from all of the weakly populated 1$^{+}$ states above 5.93~MeV excitation energy contributes only 2.1\% of the $\gamma$-feeding intensity from the 1$^{+}$ states below 5.93~MeV. The latter was measured to be 93(7)~ppm, and we thus estimate the additional $\gamma$-ray feeding from high-lying 1$^{+}$ states to be $93(7)\text{~ppm} \times 2.1\% = 2(2)\text{~ppm}$, where we have again adopted a conservative $\pm100\%$ relative uncertainty in this theoretical estimate. The direct nonanalogue Fermi $\beta$ branch to the 0$^{+}_{1}$ level at 2342~keV excitation energy is thus determined to be $95(7) + 4(8) - 2(2)\text{~ppm} = 97(10)\text{~ppm}$. Combining this nonanalogue $\beta$ branching ratio with the ratio of phase space integrals $f_{0}/f_{1} = 4.91$, yields an isospin-mixing component $\delta_{C1}^{1}$ which is determined to be 0.048(5)\%.

The isospin mixing with the first-excited 0$^{+}$ state of $^{62}$Zn determined in the current experiment is thus seen to be a factor of 1.7 times lower than the value of $\delta_{C1}^{1} = 0.083\%$ obtained from the MSDI shell-model calculations by scaling the bare shell-model value of $\delta_{C1}^{1} = 0.089\%$ by the square of the ratio of the theoretical and experimental excitation energies for this 0$^{+}$ state $(2.263/2.342)^2$, and a factor of 3.3 smaller than the value of $\delta_{C1}^{1} = 0.158\%$ predicted by the GXPF1A shell-model calculations, again with scaling of the bare shell model value of 0.160$\%$ by the excitation energy ratio $(2.321/2.342)^{2}$. We note that the experimental $\delta_{C1}^{1} = 0.048(5)\%$ is also a factor of 1.6 smaller than the value of $\delta_{C1}^{1} = 0.079\%$ calculated with the FPVH shell-model interaction and a factor of 3.5 lower than the value of $\delta_{C1}^{1} = 0.169\%$ predicted with the FPD6* interaction in the earlier shell-model calculations of Ref.~\cite{Ormand1995}. All of the shell-model calculations performed to date for the superallowed $\beta$ decay of $^{62}$Ga are thus seen to significantly overestimate the isospin mixing with the first excited 0$^{+}$ state of $^{62}$Zn by factors that range from approximately 1.6 to 3.5.

\begin{figure}[ht!]
\begin{center}
\includegraphics[width=\columnwidth,keepaspectratio]{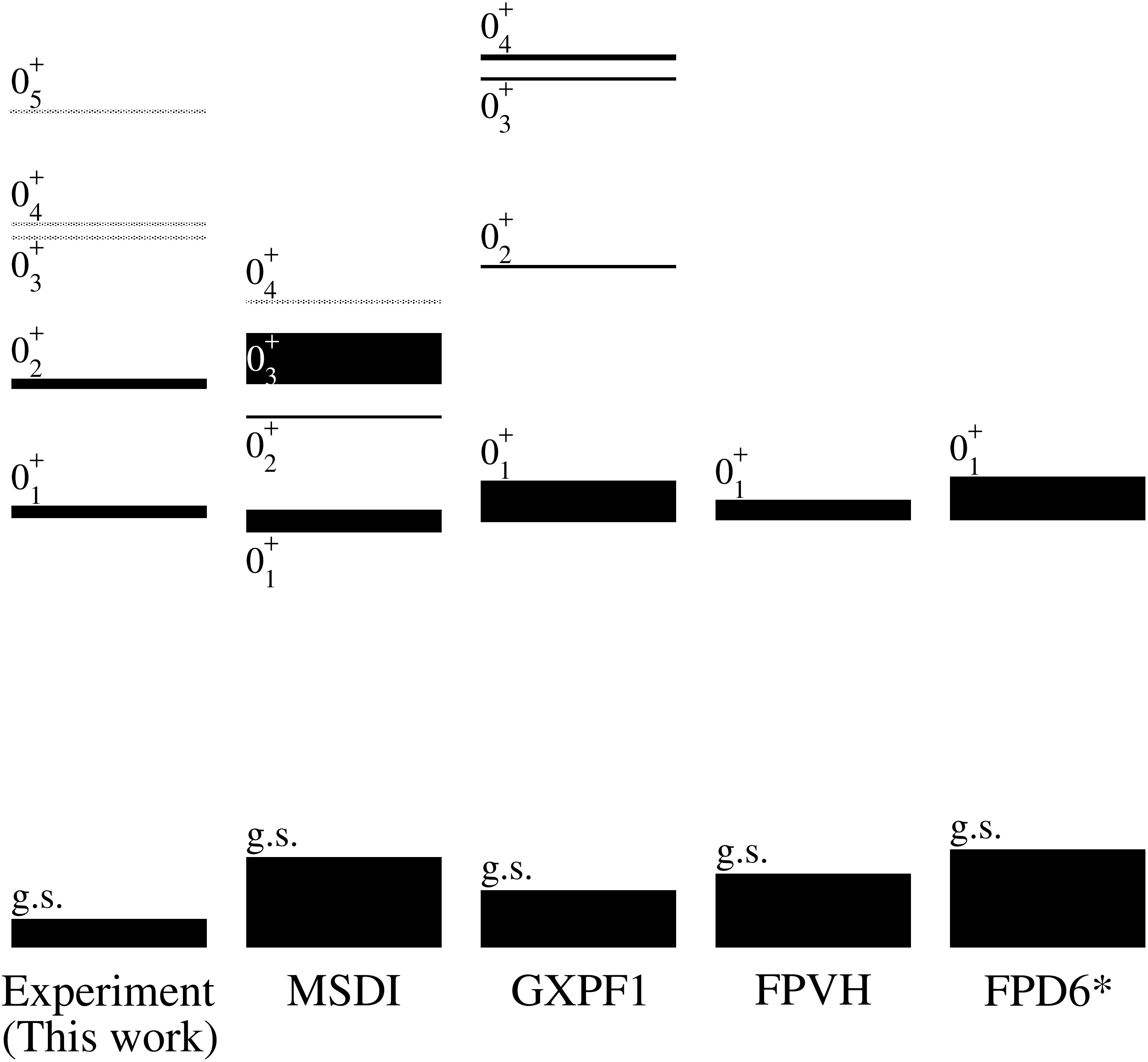}
\caption{A comparison between the current experimental results and the theoretical calculations of Refs.~\cite{Ormand1995,Towner2002,Towner2008} for the isospin mixing components with low-lying $0^{+}$ states in the daughter nucleus $^{62}$Zn during $^{62}$Ga $\beta$ decay.  For the excited $0^{+}$ states the thicknesses of the levels are proportional to the $\delta_{C1}^{n}$ values, while for the ground state the thickness is proportional to the total $\delta_{C1}$. The dashed lines in the experimental level scheme represent excited $0^{+}$ states from Ref.~\cite{Leach2013} that were not observed in the current work and for which only upper limits could be set on the $\delta_{C1}^{n}$ values.}
\label{fig:deltaC1}
\end{center}
\end{figure}

As noted, the MSDI shell-model calculation for $^{62}$Ga $\beta$ decay predicts that the primary isospin mixing is not with the first-excited 0$^{+}$ state in $^{62}$Zn but rather with the third-excited 0$^{+}$ state calculated at 3.072~MeV excitation energy with a $\delta_{C1}^{3} = 0.199\%$. Associating this theoretical prediction with the experimentally observed 0$^{+}_{2}$ state at 3.046~MeV excitation energy, the experimentally measured $I_{out}-I_{in} = 45(2)$~ppm for this state can be used to test the theoretical predictions. Following the same procedure as for the 0$^{+}_{1}$ state discussed above, limits on the intensities of possible transitions into and out of the 3.046-MeV 0$^{+}_{2}$ state involving the identified levels in $^{62}$Zn shown in Fig.~\ref{fig:LevelScheme} were again set by analyzing $\gamma-\gamma$ coincidence spectra and a possible $\gamma$ feeding from unobserved high-lying states of 2(2)~ppm where a 100$\%$ relative uncertainty was again adopted. This leads to our best estimate of the direct nonanalogue Fermi $\beta$ feeding of the 0$^{+}_{2}$ level at 3.046~MeV as $45(2) + 1.8(11) - 2(2)$~ppm $= 45(3)$~ppm. Combining this result with the ratio of phase space integrals $(f_{0}/f_{2}) = 8.93$ yields the isospin-mixing component $\delta_{C1}^{2} = 0.040(3)\%$. This result is larger than the limit $\delta_{C1}^{2} \leq 0.011(4)\%$ reported in Ref.~\cite{Finlay2008} because of the reassignment of the intensity between the two components of the 2092/2093-keV $\gamma$-ray doublet that was enabled by the $\gamma-\gamma$ coincidence analysis described in Section~\ref{sec:gamma}. The current result, $\delta_{C1}^{2} = 0.040(3)\%$, thus supersedes the limit reported in Ref.~\cite{Finlay2008}. The current result suggests that the isospin-mixing with the 0$^{+}$ state at 3.046~MeV excitation energy is comparable to the mixing with the first excited 0$^{+}_{1}$ state at 2.342~MeV excitation. The $\delta_{C1}^{2} = 0.040(3)\%$ is, however, a factor of 5 lower than the value of 0.199$\%$ predicted by the MSDI calculations for a 0$^{+}$ state predicted at 3.072~MeV excitation. 

Adding the experimental values for the isospin-mixing components with the first two excited 0$^{+}$ states observed in the current work gives $\delta_{C1}^{1} + \delta_{C1}^{2} = 0.088(6)\%$. In both the MSDI and GXPF1 calculations, mixing with two low-lying excited 0$^{+}$ states (the 0$^{+}_{1}$ and 0$^{+}_{3}$ states in the MSDI and the 0$^{+}_{1}$ and 0$^{+}_{2}$ in the GXPF1 calculations) account for $\approx 80(3)\%$ of the total isospin-mixing correction $\delta_{C1}$. The experimental sum thus suggests an isospin-mixing correction for $^{62}$Ga superallowed Fermi $\beta$ decay of $\delta_{C1} = 0.088(6)\%/0.80(3) = 0.110(9)\%$. This value is a factor of 2.6 smaller than the currently adopted correction of $\delta_{C1} = 0.275(55)\%$ and well outside of its quoted uncertainty range. The conclusion is therefore that either current theoretical models are significantly overestimating the configuration mixing component $\delta_{C1}$ of the isospin symmetry breaking correction for the superallowed $\beta$ decay of $^{62}$Ga, or this mixing is shifted to more highly-excited 0$^{+}$ states in the $^{62}$Zn daughter nucleus, a conclusion that would also be inconsistent with current theoretical models.

To further explore the latter possibility, we note that in addition to the first two excited 0$^{+}$ states in $^{62}$Zn at 2342 and 3046~keV excitation energy for which direct nonanalogue Fermi $\beta$ branches were conclusively observed in the current experiment, the $^{64}$Zn($p,t$)$^{62}$Zn transfer reaction study of Ref.~\cite{Leach2013} also reported excited 0$^{+}$ states in $^{62}$Zn at excitation energies of 3862(2), 3936(6) and 4552(9)~keV. While none of these higher-lying excited 0$^{+}$ states were observed in the current $^{62}$Ga $\beta$ decay experiment, upper limits on their nonanalogue Fermi $\beta$ feeding branches were determined by analyzing $\gamma$-ray spectra in coincidence with the 954-keV $2^{+}_{1}\rightarrow 0^{+}_{gs}$ and 1805-keV $2^{+}_{2}\rightarrow 0^{+}_{gs}$ transitions under the assumption that the $\gamma$-decay from these higher-lying excited 0$^{+}$ states would proceed primarily to the first two excited 2$^{+}$ states of $^{62}$Zn. In this analysis, peaks were fit at the expected locations of the $\gamma$-ray transitions feeding the 954-keV and 1805-keV states with peak-shape parameters fixed but the centroid energies allowed to vary within the uncertainties of the excitation energies reported in the transfer reaction work of Ref.~\cite{Leach2013}. The resulting limits on the $\gamma$-ray intensities to the 954-keV and 1805-keV states were then summed to establish the upper limits on the nonanalogue Fermi decays to these states shown in Table~\ref{tab:Limits} and Figure~\ref{fig:deltaC1}, together with the results for the 0$^{+}_{1}$ and 0$^{+}_{2}$ states discussed above. 

As shown in Table~\ref{tab:Limits}, the non-observation of the 0$^{+}_{3}$, 0$^{+}_{4}$, and 0$^{+}_{5}$ levels in the current $^{62}$Ga $\beta$ decay experiment sets stringent upper limits on the nonanalogue Fermi $\beta$ branches to these states in the $\approx$1--2~ppm range. In relation to the 0$^{+}_{5}$ level, we note that a state was identified at 4532.2(5)~keV excitation energy in the current work that was observed to $\gamma$ decay to the 2$^{+}_{1}$ state but not the 0$^{+}$ ground state (or other 0$^{+}$ states). It is possible that this is the same state as the 0$^{+}_{5}$ level reported at 4552(9)~keV excitation energy in Ref.~\cite{Leach2013}. If so, the measured $I_{out}-I_{in}$ of 3.1(12)~ppm for the 4532-keV level would correspond to an isospin mixing component $\delta_{C1}^{5} = 0.014(5)\%$ which does not differ significantly from the upper limit $\delta_{C1}^{5} \leq 0.011\%$ for the 4552(9)-keV state reported in Table~\ref{tab:Limits} and does not qualitatively alter any of the conclusions drawn below. Correcting for the phase space integral ratios ($f_{0}/f_{n}$), the nonanalogue Fermi branching ratio limits for the 0$^{+}_{3}$, 0$^{+}_{4}$, and 0$^{+}_{5}$ states correspond to the upper limits on $\delta_{C1}^{3}$, $\delta_{C1}^{4}$, and $\delta_{C1}^{5}$ shown in column five of Table~\ref{tab:Limits}. Summing these upper limits indicates that the 0$^{+}_{3-5}$ states contribute $\leq$0.018\% to $\delta_{C1}$. Combined with the measured $\delta_{C1}^{1} + \delta_{C1}^{2} = 0.088(6)\%$, this suggests a total isospin-mixing correction for $^{62}$Ga $\beta$ decay in the range of $\delta_{C1} = 0.082$--0.115\%, qualitatively consistent with the value of $\delta_{C1} = 0.110(9)\%$ obtained above by assuming that $\delta_{C1}^{1} + \delta_{C1}^{2}$ contributes 80(3)$\%$ of the total $\delta_{C1}$.

The current high-precision study of $^{62}$Ga $\beta$ decay thus leads to the conclusion that the isospin-mixing component $\delta_{C1}$ of the isospin symmetry breaking correction for $^{62}$Ga superallowed Fermi $\beta$ decay is a factor of approximately 2.5 smaller than the currently adopted value of $\delta_{C1} = 0.275(55)\%$ and that all shell-model calculations for $^{62}$Ga decay performed to date are significantly overestimating the amount of configuration mixing induced by isospin-symmetry breaking effects. In other words, the $^{62}$Ga and $^{62}$Zn ground states are, in fact, closer to being perfect isobaric analogue configurations than current calculations predict. A common feature of these shell-model calculations is that they have all been performed assuming a closed $^{56}$Ni doubly-magic core, a known oversimplification to the structure of nuclei in this mass region~\cite{Honma2004}. Improved calculations of isospin-symmetry breaking effects in the $A\geq62$ superallowed Fermi $\beta$ decays will require the lifting of this approximation while also retaining the constraints on the isospin non-conserving components of the shell-model interaction obtained by requiring that they reproduce the coefficients of the Isobaric Mass Multiplet Equation (IMME) in the region, as developed in Refs.~\cite{Ormand1989, Towner2002}.


\section{Conclusion}

A high-precision measurement of the superallowed Fermi $\beta$ branching ratio in the decay of $^{62}$Ga was performed with the GRIFFIN $\gamma$-ray spectrometer at the TRIUMF-ISAC facility. The result, $B_{0} = 99.8577^{+0.0023}_{-0.0029}\%$, is consistent with, but a factor of 4 more precise than the previously adopted world-average for this quantity. The significant gain in precision was achieved through a careful control of the so-called Pandemonium effect via high-efficiency, high-resolution $\gamma$-ray spectroscopy. The high $\gamma-\gamma$ coincidence efficiency also enabled an angular correlation analysis for a $\gamma-\gamma$ cascade populated in only 0.0188$\%$ of $^{62}$Ga $\beta$ decays. This analysis firmly establishes the first excited 0$^{+}$ state in the daughter nucleus $^{62}$Zn at 2342.1(1)~keV excitation energy, resolving a discrepancy between previous studies~\cite{Albers2010, Leach2013, Kusakari1972, Hinrichs1974, Leach2019, Fulbright1977} with important implications for the scaling of theoretical isospin symmetry breaking corrections in $^{62}$Ga decay. Weak nonanalogue Fermi $\beta$ decay branches of 97(10)~ppm and 45(3)~ppm to the first two excited 0$^{+}$ states of $^{62}$Zn were measured, corresponding to isospin-mixing components $\delta_{C1}^{1} = 0.048(5)\%$ and $\delta_{C1}^{2} = 0.040(3)\%$. Combined with stringent limits set on the nonanalogue Fermi $\beta$ branches to three more highly excited 0$^{+}$ states in $^{62}$Zn, these results suggest a total isospin-mixing correction in $^{62}$Ga superallowed decay of $\delta_{C1} = 0.110(9)\%$, approximately a factor of 2.5 smaller than the currently adopted theoretical value of 0.275(55)\%~\cite{Hardy2015}. The precise measurements of, and upper limits set on, the $\delta_{C1}^{n}$ values in the current work will provide benchmarks for future improvements to the theoretical models of isospin symmetry breaking effects in high-$Z$ superallowed Fermi $\beta$ decays. 

{\it Note added in proof.} A new survey of the world superallowed Fermi $\beta$ decay data has recently been published~\cite{Hardy2020}. This survey adopted a transition-independent ``inner" radiative correction $\Delta_{R}^{V} = 2.454(19)$\% from a weighted average of the two recent calculations reported in Refs.~\cite{SengPRL2019} and \cite{CzarneckiPRD2019}, and obtained $|V_{ud}| = 0.97373(31)$ for the up-down element of the CKM quark-mixing matrix.  The larger uncertainty compared to Eq.~(3) of the current manuscript results from the additional contributions to the nuclear-structure dependent ``outer" radiative corrections, $\delta_{NS}$, described in Refs.~\cite{SengPRD2019} and \cite{Gorchtein2019}.

\begin{acknowledgments}
We would like to thank the operations and beam delivery staff at TRIUMF for providing the high-quality $^{62}$Ga radioactive beam, and I.S. Towner for providing the shell-model calculations of $^{62}$Ga $\beta$ decay and the subsequent $\gamma$ decay in the daughter nucleus $^{62}$Zn. This work was supported by the Natural Sciences and Engineering Research Council of Canada and the U.S. DOE Office of Science, under grant DE-SC0017649. The GRIFFIN infrastructure was funded jointly by the Canada Foundation for Innovation, the Ontario Ministry of Research and Innovation, the British Columbia Knowledge Development Fund, TRIUMF and the University of Guelph. TRIUMF receives funding through a contribution agreement with the National Research Council Canada. C.E.S. acknowledges support from the Canada Research Chairs program. 
\end{acknowledgments}

\bibliography{62GaPRC}
\end{document}